\documentclass[showpacs,epsf,floats,pre,onecolumn,10pt]{revtex4}
\usepackage{amssymb}
\usepackage{amsfonts}
\usepackage{amsmath}

\input{tcilatex}

\begin{document}

\title{Can chaotic quantum energy levels statistics be characterized using
information geometry and inference methods?}
\author{C. Cafaro}
\email{carlocafaro2000@yahoo.it}
\author{S. A. Ali}
\email{alis@alum.rpi.edu}
\affiliation{Department of Physics, State University of New York at Albany-SUNY,1400
Washington Avenue, Albany, NY 12222, USA}

\begin{abstract}
In this paper, we review our novel information geometrodynamical approach to
chaos (IGAC) on curved statistical manifolds and we emphasize the usefulness
of our information-geometrodynamical entropy (IGE) as an indicator of
chaoticity in a simple application. Furthermore, knowing that integrable and
chaotic quantum antiferromagnetic Ising chains are characterized by
asymptotic logarithmic and linear growths of their operator space
entanglement entropies, respectively, we apply our IGAC to present an
alternative characterization of such systems. Remarkably, we show that in
the former case the IGE exhibits asymptotic logarithmic growth while in the
latter case the IGE exhibits asymptotic linear growth.

At this stage of its development, IGAC remains an ambitious unifying\
information-geometric theoretical construct for the study of chaotic
dynamics with several unsolved problems. However, based on our recent
findings, we believe it could provide an interesting, innovative and
potentially powerful way to study and understand the very important and
challenging problems of classical and quantum chaos.\ \ \ 
\end{abstract}

\maketitle



\section{\textbf{Introduction}}

In classical and quantum dynamics there is no unified characterization of
chaos. In the Riemannian \cite{casetti} and Finslerian \cite{cipriani} (a
Finsler metric is obtained from a Riemannian metric by relaxing the
requirement that the metric be quadratic on each tangent space)
geometrodynamical approach to chaos in classical Hamiltonian systems, an
active field of research concerns the possibility of finding a rigorous
relation among the sectional curvature, the Lyapunov exponents, and the
Kolmogorov-Sinai dynamical entropy (i.e. the sum of positive Lyapunov
exponents) \cite{kawabe}. The largest Lyapunov exponent characterizes the
degree of chaoticity of a dynamical system and, if positive, it measures the
mean instability rate of nearby trajectories averaged along a sufficiently
long reference trajectory. Moreover, it is known that classical chaotic
systems are distinguished by their exponential sensitivity to initial
conditions and that the absence of this property in quantum systems has lead
to a number of different criteria being proposed for quantum chaos.
Exponential decay of fidelity, hypersensitivity to perturbation and the
Zurek-Paz quantum chaos criterion of linear von Neumann's entropy growth 
\cite{zurek} are some examples \cite{caves}. These criteria accurately
predict chaos in the classical limit, but it is not clear that they behave
the same far from the classical realm.

The present work makes use of the so-called Entropic Dynamics (ED) \cite%
{caticha1}. ED is a theoretical framework that arises from the combination
of inductive inference (Maximum relative Entropy Methods, \cite{caticha2})
and Information Geometry (Riemannian geometry applied to probability theory)
(IG) \cite{amari}. As such, ED is constructed on statistical manifolds. It
is developed to investigate the possibility that laws of physics - either
classical or quantum - might reflect laws of inference rather than laws of
nature.

This article is a follow up of a series of the authors works \cite{cafaro1,
cafaro2, cafaro3, cafaro4a, cafaro4, cafaro5, cafaro6, cafaro7}. Especially
the work presented in \cite{cafaro6} will be discussed in more detail. The
ED theoretical framework is used to explore the possibility of constructing
a unified characterization of classical and quantum chaos. The general
formalism of the IGAC is presented by investigating a system with $3l$
degrees of freedom (microstates), each one described by two pieces of
relevant information, its mean expected value and its variance (Gaussian
statistical macrostates). This leads to consider an ED model on a
non-maximally symmetric $6l$-dimensional statistical manifold $\mathcal{M}%
_{s}$. It is shown that $\mathcal{M}_{s}$ possesses a constant negative
Ricci curvature that is proportional to the number of degrees of freedom of
the system, $\mathcal{R}_{\mathcal{M}_{s}}=-3l$. It is shown that the system
explores statistical volume elements on $\mathcal{M}_{s}$ at an exponential
rate. We define an information geometrodynamical entropy (IGE) $\mathcal{S}_{%
\mathcal{M}_{s}}$\ of the system and we show it increases linearly in time
(statistical evolution parameter) and is moreover proportional to the number
of degrees of freedom of the system. The geodesics on $\mathcal{M}_{s}$ are
hyperbolic trajectories. Using the Jacobi-Levi-Civita (JLC) equation for
geodesic spread, it is shown that the Jacobi vector field intensity $J_{%
\mathcal{M}_{s}}$ diverges exponentially and is proportional to the number
of degrees of freedom of the system. Thus, $\mathcal{R}_{\mathcal{M}_{s}}$, $%
\mathcal{S}_{\mathcal{M}_{s}}$ and $J_{\mathcal{M}_{s}}$ are proportional to
the number of Gaussian-distributed microstates of the system. This
proportionality leads to conclude there is a substantial link among these
information-geometric indicators of chaoticity. We emphasize that our IGE
provides an information-geometric analog of the Zurek-Paz quantum chaos
criterion \cite{cafaro2}. As a physical application of our general
theoretical scheme that we have called the Information Geometrodynamical
Approach to Chaos (IGAC), we provide an information-geometric analogue of
quantum energy level statistics for integrable and chaotic quantum spin
chains. It is known \cite{prosen(07)} that in the integrable case, the
antiferromagnetic Ising chain is immersed in a transverse homogeneous
magnetic field and the level spacing distribution of its spectrum is the
Poisson distribution. Instead, in the chaotic case, the antiferromagnetic
Ising chain is immersed in a tilted homogeneous magnetic field and the level
spacing distribution of its Hamiltonian spectrum is the Wigner-Dyson
distribution. The antiferromagnetic Ising spin chain in external magnetic
field is one example of order-to-chaos transition in quantum many-body
context and it is used here as a demonstrating example of the conjectured
connection between the Wigner-Dyson (Poisson) statistics and nonitegrability
(integrability) in quantum mechanics. Moreover, it is known that integrable
and chaotic quantum antiferromagnetic Ising chains are characterized by
asymptotic logarithmic and linear growths of their operator space
entanglement entropies \cite{prosen(07)}, respectively.

Following the results provided by Prosen, we study the
information-geometrodynamics of a Poisson distribution coupled to an
Exponential bath (regular case) and that of a Wigner-Dyson distribution
coupled to a Gaussian bath (chaotic case). Remarkably, we show that in the
former case the IGE exhibits asymptotic logarithmic growth while in the
latter case it exhibits asymptotic linear growth.

The layout of this paper is as follows. In Section II, the general formalism
of the IGAC is applied to study a simple example, an ED Gaussian statistical
model. In Section III, the main indicators of chaoticity within our novel
theoretical construct are introduced by studying the ED Gaussian model. In
Section IV, special focus is devoted to the role of the IGE as an indicator
of temporal complexity on curved statistical manifolds. In Section V, after
presenting the basics of the IG of Poisson and Wigner-Dyson distributions,
we briefly review the conventional approach suitable to study the energy
level statistics of integrable and chaotic antiferromagnetic Ising chains
immersed in external magnetic fields. In Section VI, we present our
IGAC-based novel characterization of the quantum energy level statistics of
such Ising chains. Finally, in Section VII, we present our final remarks.

\section{Theoretical Structure of the IGAC: A simple example}

The IGAC\ arises as a theoretical framework to study chaos in informational
geodesic flows describing physical, biological or chemical \ systems. A
geodesic on a curved statistical manifold represents the maximum probability
path a complex dynamical system explores in its evolution between the
initial and the final macrostates. Each point of the geodesic is
parametrized by the macroscopic dynamical variables defining the macrostate
of the system. Furthermore, each macrostate is in a one-to-one relation with
the probability distribution representing the maximally probable description
of the system being considered. The set of macrostates forms the parameter
space while the set of probability distributions forms the statistical
manifold. The parameter space is homeomorphic to the statistical manifold.%
\textbf{\ }IGAC is the information-geometric analogue of conventional
geometrodynamical approaches \cite{casetti, cipriani} where the classical
configuration space $\Gamma _{E}$\ is being replaced by a statistical
manifold $\mathcal{M}_{S}$\ with the additional possibility of considering
chaotic dynamics arising from non conformally flat metrics (the Jacobi
metric is always conformally flat, instead). It is an information-geometric
extension of the Jacobi geometrodynamics (the geometrization of a
Hamiltonian system by transforming it to a geodesic flow \cite{jacobi}). The
reformulation of dynamics in terms of a geodesic problem allows the
application of a wide range of well-known geometrical techniques in the
investigation of the solution space and properties of the equation of
motion. The power of the Jacobi reformulation is that all of the dynamical
information is collected into a single geometric object in which all the
available manifest symmetries are retained- the manifold on which geodesic
flow is induced.

Using information-geometric methods, we have investigated in some detail the
still open problem of finding a unifying description of classical and
quantum chaos \cite{cafaro2}. One of our goals in this paper is that of
representing an additional step forward in that research direction.

\subsection{The ED\ Gaussian Model}

Maximum relative Entropy (ME) methods are used to construct an ED model that
follows from an assumption about what information is relevant to predict the
evolution of the system. Given a known initial macrostate (probability
distribution) and that the system evolves to a final known macrostate, the
possible trajectories of the system are examined. A notion of \textit{%
distance} between two probability distributions is provided by IG. As shown
in \cite{fisher, rao} this distance is quantified by the Fisher-Rao
information metric tensor.

In the following example, we consider an ED model whose microstates span a $%
3l$-dimensional space labelled by the variables $\left\{ \vec{X}\right\}
=\left\{ \vec{x}^{\left( 1\right) }\text{, }\vec{x}^{\left( 2\right) }\text{%
,...., }\vec{x}^{\left( l\right) }\right\} $ with $\vec{x}^{\left( \alpha
\right) }\equiv \left( x_{1}^{\left( \alpha \right) }\text{, }x_{2}^{\left(
\alpha \right) }\text{, }x_{3}^{\left( \alpha \right) }\right) $, $\alpha =1$%
,...., $l$ and $x_{a}^{\left( \alpha \right) }\in 
\mathbb{R}
$ with $a=1$, $2$, $3$. We assume the only testable information pertaining
to the $3l$ degrees of freedom $\left\{ x_{a}^{\left( \alpha \right)
}\right\} $ consists of the expectation values $\left\langle x_{a}^{\left(
\alpha \right) }\right\rangle $ and variances $\Delta x_{a}^{\left( \alpha
\right) }\equiv \sqrt{\left\langle \left( x_{a}^{\left( \alpha \right)
}-\left\langle x_{a}^{\left( \alpha \right) }\right\rangle \right)
^{2}\right\rangle }$. The set of these expectation values define the $6l$%
-dimensional space of macrostates of the system. A measure of
distinguishability among the macrostates of the ED model is obtained by
assigning a probability distribution $P\left( \vec{X}|\vec{\Theta}\right) $
to each macrostate $\vec{\Theta}$ where $\left\{ \vec{\Theta}\right\}
=\left\{ ^{\left( 1\right) }\theta _{a}^{\left( \alpha \right) }\text{, }%
^{\left( 2\right) }\theta _{a}^{\left( \alpha \right) }\right\} $ with $%
\alpha =1$, $2$,$....$, $l$ and $a=1$, $2$, $3$. The process of assigning a
probability distribution to each state endows $\mathcal{M}_{S}$ with a
metric structure. Specifically, the Fisher-Rao information metric is a
measure of distinguishability among macrostates. It assigns an IG to the
space of states. Each macrostate may be viewed as a point of a $6l$%
-dimensional statistical manifold with coordinates given by the numerical
values of the expectations $\left\langle x_{a}^{\left( \alpha \right)
}\right\rangle =^{\left( 1\right) }\theta _{a}^{\left( \alpha \right) }$ and 
$\Delta x_{a}^{\left( \alpha \right) }\equiv \sqrt{\left\langle \left(
x_{a}^{\left( \alpha \right) }-\left\langle x_{a}^{\left( \alpha \right)
}\right\rangle \right) ^{2}\right\rangle }=$ $^{\left( 2\right) }\theta
_{a}^{\left( \alpha \right) }$. The available information can be written in
the form of the following $6l$ information constraint equations,%
\begin{equation}
\begin{array}{c}
\left\langle x_{a}^{\left( \alpha \right) }\right\rangle
=\dint\limits_{-\infty }^{+\infty }dx_{a}^{\left( \alpha \right)
}x_{a}^{\left( \alpha \right) }P_{a}^{\left( \alpha \right) }\left(
x_{a}^{\left( \alpha \right) }\left\vert ^{\left( 1\right) }\theta
_{a}^{\left( \alpha \right) }\text{,}^{\left( 2\right) }\theta _{a}^{\left(
\alpha \right) }\right. \right) \text{,} \\ 
\\ 
\Delta x_{a}^{\left( \alpha \right) }=\left[ \dint\limits_{-\infty
}^{+\infty }dx_{a}^{\left( \alpha \right) }\left( x_{a}^{\left( \alpha
\right) }-\left\langle x_{a}^{\left( \alpha \right) }\right\rangle \right)
^{2}P_{a}^{\left( \alpha \right) }\left( x_{a}^{\left( \alpha \right)
}\left\vert ^{\left( 1\right) }\theta _{a}^{\left( \alpha \right) }\text{,}%
^{\left( 2\right) }\theta _{a}^{\left( \alpha \right) }\right. \right) %
\right] ^{\frac{1}{2}}\text{.}%
\end{array}
\label{constraint1}
\end{equation}%
The probability distributions $P_{a}^{\left( \alpha \right) }$ in (\ref%
{constraint1}) are constrained by the conditions of normalization,%
\begin{equation}
\dint\limits_{-\infty }^{+\infty }dx_{a}^{\left( \alpha \right)
}P_{a}^{\left( \alpha \right) }\left( x_{a}^{\left( \alpha \right)
}\left\vert ^{\left( 1\right) }\theta _{a}^{\left( \alpha \right) }\text{,}%
^{\left( 2\right) }\theta _{a}^{\left( \alpha \right) }\right. \right) =1%
\text{.}  \label{constraint2}
\end{equation}%
Information theory identifies the Gaussian distribution as the maximum
entropy distribution if only the expectation value and the variance are
known \cite{jaynes2}. ME methods allow us to associate a probability
distribution $P\left( \vec{X}|\vec{\Theta}\right) $ to each point in the
space of states $\vec{\Theta}$ \cite{caticha2}. The distribution that best
reflects the information contained in the prior distribution $m\left( \vec{X}%
\right) $ updated by the information $\left( \left\langle x_{a}^{\left(
\alpha \right) }\right\rangle \text{, }\Delta x_{a}^{\left( \alpha \right)
}\right) $ is obtained by maximizing the relative entropy 
\begin{equation}
S\left( \vec{\Theta}\right) =-\int\limits_{\left\{ \vec{X}\right\} }d^{3l}%
\vec{X}P\left( \vec{X}\left\vert \vec{\Theta}\right. \right) \log \left( 
\frac{P\left( \vec{X}\left\vert \vec{\Theta}\right. \right) }{m\left( \vec{X}%
\right) }\right) \text{.}  \label{entropy}
\end{equation}%
As a working hypothesis, the prior $m\left( \vec{X}\right) $ is set to be
uniform since we assume the lack of prior available information about the
system (postulate of equal \textit{a priori} probabilities). Upon maximizing
(\ref{entropy}), given the constraints (\ref{constraint1}) and (\ref%
{constraint2}), we obtain%
\begin{equation}
P\left( \vec{X}\left\vert \vec{\Theta}\right. \right) =\dprod\limits_{\alpha
=1}^{l}\dprod\limits_{a=1}^{3}P_{a}^{\left( \alpha \right) }\left(
x_{a}^{\left( \alpha \right) }\left\vert \mu _{a}^{\left( \alpha \right) }%
\text{, }\sigma _{a}^{\left( \alpha \right) }\right. \right)  \label{prob}
\end{equation}%
where%
\begin{equation}
P_{a}^{\left( \alpha \right) }\left( x_{a}^{\left( \alpha \right)
}\left\vert \mu _{a}^{\left( \alpha \right) }\text{, }\sigma _{a}^{\left(
\alpha \right) }\right. \right) =\left( 2\pi \left[ \sigma _{a}^{\left(
\alpha \right) }\right] ^{2}\right) ^{-\frac{1}{2}}\exp \left[ -\frac{\left(
x_{a}^{\left( \alpha \right) }-\mu _{a}^{\left( \alpha \right) }\right) ^{2}%
}{2\left( \sigma _{a}^{\left( \alpha \right) }\right) ^{2}}\right]
\label{gaussian}
\end{equation}%
and, in the standard notation for Gaussians, $^{\left( 1\right) }\theta
_{a}^{\left( \alpha \right) }\overset{\text{def}}{=}\left\langle
x_{a}^{\left( \alpha \right) }\right\rangle \equiv \mu _{a}^{\left( \alpha
\right) }$, $^{\left( 2\right) }\theta _{a}^{\left( \alpha \right) }\overset{%
\text{def}}{=}\Delta x_{a}^{\left( \alpha \right) }\equiv \sigma
_{a}^{\left( \alpha \right) }$. The probability distribution (\ref{prob})
encodes the available information concerning the system. Note we assumed
uncoupled constraints among microvariables $x_{a}^{\left( \alpha \right) }$.
In other words, we assumed that information about correlations between the
microvariables need not to be tracked. This assumption leads to the
simplified product rule (\ref{prob}). However, coupled constraints would
lead to a generalized product rule in (\ref{prob}) and to an information
metric tensor with non-trivial off-diagonal elements (covariance terms). For
instance, the total probability distribution $P\left( x\text{, }y|\mu _{x}%
\text{, }\sigma _{x}\text{, }\mu _{y}\text{, }\sigma _{y}\right) $ of two
dependent Gaussian distributed microvariables $x$ and $y$ reads%
\begin{eqnarray}
&&P\left( x\text{, }y|\mu _{x}\text{, }\sigma _{x}\text{, }\mu _{y}\text{, }%
\sigma _{y}\right) =\frac{1}{2\pi \sigma _{x}\sigma _{y}\sqrt{1-r^{2}}}\times
\label{corr-prob} \\
&&\times \exp \left\{ -\frac{1}{2\left( 1-r^{2}\right) }\left[ \frac{\left(
x-\mu _{x}\right) ^{2}}{\sigma _{x}^{2}}-2r\frac{\left( x-\mu _{x}\right)
\left( y-\mu _{y}\right) }{\sigma _{x}\sigma _{y}}+\frac{\left( y-\mu
_{y}\right) ^{2}}{\sigma _{y}^{2}}\right] \right\} \text{,}  \notag
\end{eqnarray}%
where $r\in \left( -1\text{, }+1\right) $ is the correlation coefficient
given by%
\begin{equation}
r=\frac{\left\langle \left( x-\left\langle x\right\rangle \right) \left(
y-\left\langle y\right\rangle \right) \right\rangle }{\sqrt{\left\langle
x-\left\langle x\right\rangle \right\rangle }\sqrt{\left\langle
y-\left\langle y\right\rangle \right\rangle }}=\frac{\left\langle
xy\right\rangle -\left\langle x\right\rangle \left\langle y\right\rangle }{%
\sigma _{x}\sigma _{y}}\text{.}
\end{equation}%
The information metric tensor induced by (\ref{corr-prob}) is \cite{cafaro7},%
\begin{equation}
g_{ij}=\left[ 
\begin{array}{cccc}
-\frac{1}{\sigma _{x}^{2}\left( r^{2}-1\right) } & 0 & \frac{r}{\sigma
_{x}\sigma _{y}\left( r^{2}-1\right) } & 0 \\ 
0 & -\frac{2-r^{2}}{\sigma _{x}^{2}\left( r^{2}-1\right) } & 0 & \frac{r^{2}%
}{\sigma _{x}\sigma _{y}\left( r^{2}-1\right) } \\ 
\frac{r}{\sigma _{x}\sigma _{y}\left( r^{2}-1\right) } & 0 & -\frac{1}{%
\sigma _{y}^{2}\left( r^{2}-1\right) } & 0 \\ 
0 & \frac{r^{2}}{\sigma _{x}\sigma _{y}\left( r^{2}-1\right) } & 0 & -\frac{%
2-r^{2}}{\sigma _{y}^{2}\left( r^{2}-1\right) }%
\end{array}%
\right] \text{,}  \label{corr-metric}
\end{equation}%
where $i$, $j=1$, $2$, $3$, $4$. The Ricci curvature scalar associated with
manifold characterized by (\ref{corr-metric}) is given by%
\begin{equation}
R=g^{ij}R_{ij}=-\frac{8\left( r^{2}-2\right) +2r^{2}\left( 3r^{2}-2\right) }{%
8\left( r^{2}-1\right) }\text{.}
\end{equation}%
It is clear that in the limit $r\rightarrow 0$, the off-diagonal elements of 
$g_{ij}$ vanish and the scalar $\mathcal{R}$ reduces to the result obtained
in \cite{cafaro2}, namely $\mathcal{R}=-2<0$. We could have in principle
considered a correlated Gaussian process characterized by two correlated
Gaussian microvariables\textbf{\ }$x$\textbf{\ }and\textbf{\ }$y$. Such a
process would lead to an ED on a five-dimensional statistical manifold whose
elements are probability distributions of the form\textbf{\ }$P\left( x\text{%
, }y|\mu _{x}\text{, }\sigma _{x}\text{, }\mu _{y}\text{, }\sigma _{y}\text{%
, }\sigma _{xy}\right) $\textbf{\ }coordinatized by the expectation values%
\textbf{\ }$\mu _{x}$\textbf{\ }and\textbf{\ }$\mu _{y}$\textbf{\ }as well
as the square-root of the three independent elements of the symmetric
covariance matrix, namely, $\sigma _{x}$,\textbf{\ }$\sigma _{y}$\textbf{\ }%
and\textbf{\ }$\sigma _{xy}$. However, in view of the computational
difficulty in obtaining analytical expressions for the elements of a\textbf{%
\ }$5\times 5$\textbf{\ }information metric with\textbf{\ }$\sigma _{xy}$%
\textbf{\ }playing the role of a macro-dynamical variable, we have chosen to
consider the ED on a four-dimensional statistical manifold whose elements
are given in (\ref{corr-prob}).\textbf{\ }Correlation terms may be
fictitious. They may arise for instance from coordinate transformations. On
the other hand, correlations may arise from external fields in which the
system is immersed. In such situations, correlations among $x_{a}^{\left(
\alpha \right) }$ effectively describe interaction between the
microvariables and the external fields. Such generalizations would require
more delicate analysis.

We cannot determine the evolution of microstates of the system since the
available information is insufficient. Not only is the information available
insufficient but we also do not know the equation of motion. In fact there
is no standard "equation of motion".\ Instead we can ask: how close are the
two total distributions with parameters $(\mu _{a}^{\left( \alpha \right) }$%
, $\sigma _{a}^{\left( \alpha \right) })$ and $(\mu _{a}^{\left( \alpha
\right) }+d\mu _{a}^{\left( \alpha \right) }$, $\sigma _{a}^{\left( \alpha
\right) }+d\sigma _{a}^{\left( \alpha \right) })$? Once the states of the
system have been defined, the next step concerns the problem of quantifying
the notion of change from the macrostate $\vec{\Theta}$ to the macrostate $%
\vec{\Theta}+d\vec{\Theta}$. A convenient measure of change is distance. The
measure we seek is given by the dimensionless \textit{distance} $ds$ between 
$P\left( \vec{X}\left\vert \vec{\Theta}\right. \right) $ and $P\left( \vec{X}%
\left\vert \vec{\Theta}+d\vec{\Theta}\right. \right) $,%
\begin{equation}
ds^{2}=g_{\mu \nu }d\Theta ^{\mu }d\Theta ^{\nu }\text{ with }\mu \text{, }%
\nu =1\text{, }2\text{,.., }6l  \label{line-element}
\end{equation}%
where%
\begin{equation}
g_{\mu \nu }=\int d\vec{X}P\left( \vec{X}\left\vert \vec{\Theta}\right.
\right) \frac{\partial \log P\left( \vec{X}\left\vert \vec{\Theta}\right.
\right) }{\partial \Theta ^{\mu }}\frac{\partial \log P\left( \vec{X}%
\left\vert \vec{\Theta}\right. \right) }{\partial \Theta ^{\nu }}
\label{fisher-rao}
\end{equation}%
is the Fisher-Rao information metric. Substituting (\ref{prob}) into (\ref%
{fisher-rao}), the metric $g_{\mu \nu }$ on $\mathcal{M}_{s}$ becomes a $%
6l\times 6l$ matrix $M$ made up of $3l$ blocks $M_{2\times 2}$ with
dimension $2\times 2$ given by,%
\begin{equation}
M_{2\times 2}=\left( 
\begin{array}{cc}
\left( \sigma _{a}^{\left( \alpha \right) }\right) ^{-2} & 0 \\ 
0 & 2\times \left( \sigma _{a}^{\left( \alpha \right) }\right) ^{-2}%
\end{array}%
\right)  \label{m-matrix}
\end{equation}%
with $\alpha =1$, $2$,$....$, $l$ and $a=1,2,3$. From (\ref{fisher-rao}),
the "length" element (\ref{line-element}) reads,%
\begin{equation}
ds^{2}=\dsum\limits_{\alpha =1}^{l}\dsum\limits_{a=1}^{3}\left[ \frac{1}{%
\left( \sigma _{a}^{\left( \alpha \right) }\right) ^{2}}d\mu _{a}^{\left(
\alpha \right) 2}+\frac{2}{\left( \sigma _{a}^{\left( \alpha \right)
}\right) ^{2}}d\sigma _{a}^{\left( \alpha \right) 2}\right] \text{.}
\end{equation}%
We bring attention to the fact that the metric structure of $\mathcal{M}_{s}$
is an emergent (not fundamental) structure. It arises only after assigning a
probability distribution $P\left( \vec{X}\left\vert \vec{\Theta}\right.
\right) $ to each state $\vec{\Theta}$.

\section{Information-Geometric Indicators of Chaos within the IGAC}

The relevant indicators of chaoticity within the IGAC\ are the Ricci scalar
curvature $\mathcal{R}_{\mathcal{M}_{s}}$\ (or, more correctly, the
sectional curvature $\mathcal{K}_{\mathcal{M}_{S}}$), the Jacobi vector
field intensity $J_{\mathcal{M}_{S}}$\ and the IGE $\mathcal{S}_{\mathcal{M}%
_{s}}$ once the line element on the curved statistical manifold $\mathcal{M}%
_{s}$\ underlying the entropic dynamics has been specified.

\subsection{Ricci Scalar Curvature, Anisotropy and Compactness}

Given the Fisher-Rao information metric, we use standard differential
geometry methods applied to the space of probability distributions to
characterize the geometric properties of $\mathcal{M}_{s}$. Recall that the
Ricci scalar curvature $\mathcal{R}$ is given by,%
\begin{equation}
\mathcal{R}=g^{\mu \nu }R_{\mu \nu }\text{,}  \label{ricci-scalar}
\end{equation}%
where $g^{\mu \nu }g_{\nu \rho }=\delta _{\rho }^{\mu }$ so that $g^{\mu \nu
}=\left( g_{\mu \nu }\right) ^{-1}$. The Ricci tensor $R_{\mu \nu }$ is
given by,%
\begin{equation}
R_{\mu \nu }=\partial _{\gamma }\Gamma _{\mu \nu }^{\gamma }-\partial _{\nu
}\Gamma _{\mu \lambda }^{\lambda }+\Gamma _{\mu \nu }^{\gamma }\Gamma
_{\gamma \eta }^{\eta }-\Gamma _{\mu \gamma }^{\eta }\Gamma _{\nu \eta
}^{\gamma }\text{.}  \label{ricci-tensor}
\end{equation}%
The Christoffel symbols $\Gamma _{\mu \nu }^{\rho }$ appearing in the Ricci
tensor are defined in the standard manner as, 
\begin{equation}
\Gamma _{\mu \nu }^{\rho }=\frac{1}{2}g^{\rho \sigma }\left( \partial _{\mu
}g_{\sigma \nu }+\partial _{\nu }g_{\mu \sigma }-\partial _{\sigma }g_{\mu
\nu }\right) .  \label{connection}
\end{equation}%
Using (\ref{m-matrix}) and the definitions given above, we can show that the
Ricci scalar curvature becomes%
\begin{equation}
R_{\mathcal{M}_{s}}=R_{\text{ }\alpha }^{\alpha }=\sum_{\rho \neq \sigma
}K\left( e_{\rho }\text{, }e_{\sigma }\right) =-3l<0\text{.}  \label{Ricci}
\end{equation}%
The scalar curvature is the sum of all sectional curvatures of planes
spanned by pairs of orthonormal basis elements $\left\{ e_{\rho }=\partial
_{\Theta _{\rho }(p)}\right\} $ of the tangent space $T_{p}\mathcal{M}_{s}$
with $p\in \mathcal{M}_{s}$, 
\begin{equation}
\mathcal{K}\left( a\text{, }b\right) =\frac{R_{\mu \nu \rho \sigma }a^{\mu
}b^{\nu }a^{\rho }b^{\sigma }}{\left( g_{\mu \sigma }g_{\nu \rho }-g_{\mu
\rho }g_{\nu \sigma }\right) a^{\mu }b^{\nu }a^{\rho }b^{\sigma }}\text{, }%
a=\sum_{\rho }\left\langle a\text{, }h^{\rho }\right\rangle e_{\rho }\text{,}
\label{sectionK}
\end{equation}%
where $\left\langle e_{\rho }\text{, }h^{\sigma }\right\rangle =\delta
_{\rho }^{\sigma }$. Notice that the sectional curvatures completely
determine the curvature tensor. From (\ref{Ricci}) we conclude that $%
\mathcal{M}_{s}$ is a $6l$-dimensional statistical manifold of constant
negative Ricci scalar curvature. A detailed analysis on the calculation of
Christoffel connection coefficients using the ED formalism for a
four-dimensional manifold of Gaussians can be found in \cite{cafaro2}.

It can be shown that $\mathcal{M}_{s}$ is not a pseudosphere (maximally
symmetric manifold). The first way this can be understood is from the fact
that the Weyl Projective curvature tensor \cite{goldberg} (or the anisotropy
tensor) $W_{\mu \nu \rho \sigma }$ defined by%
\begin{equation}
W_{\mu \nu \rho \sigma }=R_{\mu \nu \rho \sigma }-\frac{R_{\mathcal{M}_{s}}}{%
n\left( n-1\right) }\left( g_{\nu \sigma }g_{\mu \rho }-g_{\nu \rho }g_{\mu
\sigma }\right) \text{,}  \label{Weyl}
\end{equation}%
with $n=6l$ in the present case, is non-vanishing. In (\ref{Weyl}), the
quantity $R_{\mu \nu \rho \sigma }$ is the Riemann curvature tensor defined
in the usual manner by%
\begin{equation}
R^{\alpha }\,_{\beta \rho \sigma }=\partial _{\sigma }\Gamma _{\text{ \ }%
\beta \rho }^{\alpha }-\partial _{\rho }\Gamma _{\text{ \ }\beta \sigma
}^{\alpha }+\Gamma ^{\alpha }\,_{\lambda \sigma }\Gamma ^{\lambda }\,_{\beta
\rho }-\Gamma ^{\alpha }\,_{\lambda \rho }\Gamma ^{\lambda }\,_{\beta \sigma
}\text{.}
\end{equation}%
Considerations regarding the negativity of the Ricci curvature as a strong
criterion of dynamical instability and the necessity of compactness of $%
\mathcal{M}_{s}$\textit{\ }in "true" chaotic dynamical systems would require
additional investigation.

The issue of symmetry of $\mathcal{M}_{s}$ can alternatively be understood
from consideration of the sectional curvature. In view of (\ref{sectionK}),
the negativity of the Ricci scalar implies the existence of expanding
directions in the configuration space manifold $\mathcal{M}_{s}$. Indeed,
from (\ref{Ricci}) one may conclude that negative principal curvatures
(extrema of sectional curvatures) dominate over positive ones. Thus, the
negativity of the Ricci scalar is only a \textit{sufficient} (not necessary)
condition for local instability of geodesic flow. For this reason, the
negativity of the scalar provides a \textit{strong }criterion of local
instability. Scenarios may arise where negative sectional curvatures are
present, but the positive ones could prevail in the sum so that the Ricci
scalar is non-negative despite the instability in the flow in those
directions. Consequently, the signs of the sectional curvatures are of
primary significance for the proper characterization of chaos.

Yet another useful way to understand the anisotropy of the $\mathcal{M}_{s}$
is the following. It is known that in $n$ dimensions, there are at most $%
\frac{n\left( n+1\right) }{2}$ independent Killing vectors (directions of
symmetry of the manifold). Since $\mathcal{M}_{s}$ is not a pseudosphere,
the information metric tensor does not admit the maximum number of Killing
vectors $K_{\nu }$ defined as%
\begin{equation}
\mathcal{L}_{K}g_{\mu \nu }=D_{\mu }K_{\nu }+D_{\nu }K_{\mu }=0\text{,}
\end{equation}%
where $D_{\mu }$, defined as%
\begin{equation}
D_{\mu }K_{\nu }=\partial _{\mu }K_{\nu }-\Gamma _{\nu \mu }^{\rho }K_{\rho }
\end{equation}%
is the covariant derivative operator with respect to the connection $\Gamma $
defined in (\ref{connection}). The Lie derivative $\mathcal{L}_{K}g_{\mu \nu
}$ of the tensor field $g_{\mu \nu }$ along a given direction $K$ measures
the intrinsic variation of the field along that direction (that is, the
metric tensor is Lie transported along the Killing vector) \cite{clarke}.
Locally, a maximally symmetric space of Euclidean signature is either a
plane, a sphere, or a hyperboloid, depending on the sign of $R$. In our
case, none of these scenarios occur. As will be seen in what follows, this
fact has a significant impact on the integration of the geodesic deviation
equation on $\mathcal{M}_{s}$. At this juncture, we emphasize it is known
that the anisotropy of the manifold underlying system dynamics plays a
crucial role in the mechanism of instability. In particular, fluctuating
sectional curvatures require also that the manifold be anisotropic. However,
the connection between curvature variations along geodesics and anisotropy
is far from clear and is currently under investigation.

Krylov was the first to emphasize \cite{krylov} the use of $\mathcal{R}<0$
as an instability criterion in the context of an $N$-body system (a gas)
interacting via Van der Waals forces, with the ultimate hope to understand
the relaxation process in a gas. However, Krylov neglected the problem of
compactness of the configuration space manifold which is important for
making inferences about exponential mixing of geodesic flows \cite{pellicott}%
. Why is compactness so significant in the characterization of chaos? 
\textit{True} chaos should be identified by the occurrence of two crucial
features: 1) strong dependence on initial conditions and exponential
divergence of the Jacobi vector field intensity, i.e., stretching of
dynamical trajectories; 2) compactness of the configuration space manifold,
i.e., folding of dynamical trajectories. Compactness \cite{cipriani, jost}
is required in order to discard trivial exponential growths due to the
unboundedness of the "volume" available to the dynamical system. In other
words, the folding is necessary to have a dynamics actually able to mix the
trajectories, making practically impossible, after a finite interval of
time, to discriminate between trajectories which were very nearby each other
at the initial time. When the space is not compact, even in presence of
strong dependence on initial conditions, it could be possible in some
instances (though not always), to distinguish among different trajectories
originating within a small distance and then evolved subject to exponential
instability.

As a final remark, we emphasize that it is known from IG \cite{amari} that
there is a one-to-one relation between elements of the statistical manifold
and the parameter space. More precisely, the statistical manifold $\mathcal{M%
}_{s}$ is homeomorphic to the parameter space $\mathcal{D}_{\Theta }$. This
implies the existence of a continuous, bijective map $h_{\mathcal{M}_{s}%
\text{, }\mathcal{D}_{\Theta }}$,%
\begin{equation}
h_{\mathcal{M}_{s}\text{, }\mathcal{D}_{\Theta }\text{ }}:\mathcal{M}_{S}\ni
P\left( \vec{X}\left\vert \vec{\Theta}\right. \right) \rightarrow \vec{\Theta%
}\in \mathcal{D}_{\Theta }
\end{equation}%
where $h_{\mathcal{M}_{s}\text{, }\mathcal{D}_{\Theta }\text{ }}^{-1}\left( 
\vec{\Theta}\right) =P\left( \vec{X}\left\vert \vec{\Theta}\right. \right) $%
. The inverse image $h_{\mathcal{M}_{s}\text{, }\mathcal{D}_{\Theta }\text{ }%
}^{-1}$ is the so-called homeomorphism map. In addition, since
homeomorphisms preserve compactness, it is sufficient to restrict ourselves
to a compact subspace of the parameter space $\mathcal{D}_{\Theta }$ in
order to ensure that $\mathcal{M}_{S}$ is itself compact.

\subsection{Canonical Formalism}

The geometrization of a Hamiltonian system by transforming it to a geodesic
flow is a well-known technique of classical mechanics associated with the
name of Jacobi \cite{jacobi}. Transformation to geodesic motion is obtained
in two steps: 1) conformal transformation of the metric; 2) rescaling of the
time parameter \cite{biesiada1}. The reformulation of dynamics in terms of a
geodesic problem allows the application of a wide range of well-known
geometrical techniques in the investigation of the solution space and
properties of equations of motions. The power of the Jacobi reformulation is
that all of the dynamical information is collected into a single geometric
object - the manifold on which geodesic flow is induced - in which all the
available manifest symmetries are retained. For instance, integrability of
the system is connected with the existence of Killing vectors and tensors on
this manifold \cite{biesiada2, uggla}.

In this Section we study the trajectories of the system on $\mathcal{M}_{s}$%
. We emphasize ED can be derived from a standard principle of least action
(of Maupertuis-Euler-Lagrange-Jacobi type) \cite{caticha1, arnold}. The main
differences are that the dynamics being considered here, namely ED, is
defined on a space of probability distributions $\mathcal{M}_{s}$, not on an
ordinary linear space $V$ and the standard coordinates $q_{\mu }$ of the
system are replaced by statistical macrovariables $\Theta ^{\mu }$. The
geodesic equations for the macrovariables of the Gaussian ED model are given
by,%
\begin{equation}
\frac{d^{2}\Theta ^{\mu }}{d\tau ^{2}}+\Gamma _{\nu \rho }^{\mu }\frac{%
d\Theta ^{\nu }}{d\tau }\frac{d\Theta ^{\rho }}{d\tau }=0  \label{geodesic}
\end{equation}%
with $\mu =1$, $2$,..., $6l$. Observe the geodesic equations are\textit{\ }%
nonlinear second order coupled ordinary differential equations. They
describe a reversible dynamics whose solution is the trajectory between an
initial and a final macrostate. The trajectory can be equally well traversed
in both directions.

\subsubsection{Geodesics on $\mathcal{M}_{s}$}

We determine the explicit form of (\ref{geodesic}) for the pairs of
statistical coordinates $(\mu _{a}^{\left( \alpha \right) }$, $\sigma
_{a}^{\left( \alpha \right) })$. Substituting the expression of the
Christoffel connection coefficients into (\ref{geodesic}), the geodesic
equations for the macrovariables $\mu _{a}^{\left( \alpha \right) }$ and $%
\sigma _{a}^{\left( \alpha \right) }$ associated to the microstate $%
x_{a}^{\left( \alpha \right) }$ become,%
\begin{equation}
\frac{d^{2}\mu _{a}^{\left( \alpha \right) }}{d\tau ^{2}}-\frac{2}{\sigma
_{a}^{\left( \alpha \right) }}\frac{d\mu _{a}^{\left( \alpha \right) }}{%
d\tau }\frac{d\sigma _{a}^{\left( \alpha \right) }}{d\tau }=0\text{, }\frac{%
d^{2}\sigma _{a}^{\left( \alpha \right) }}{d\tau ^{2}}-\frac{1}{\sigma
_{a}^{\left( \alpha \right) }}\left( \frac{d\sigma _{a}^{\left( \alpha
\right) }}{d\tau }\right) ^{2}+\frac{1}{2\sigma _{a}^{\left( \alpha \right) }%
}\left( \frac{d\mu _{a}^{\left( \alpha \right) }}{d\tau }\right) ^{2}=0\text{%
,}
\end{equation}%
with $\alpha =1$, $2$,$....$, $l$ and $a=1$, $2$, $3$. This is a set of
coupled ordinary differential equations, whose solutions are%
\begin{equation}
\begin{array}{c}
\mu _{a}^{\left( \alpha \right) }\left( \tau \right) =\frac{\frac{\left(
B_{a}^{\left( \alpha \right) }\right) ^{2}}{2\beta _{a}^{\left( \alpha
\right) }}}{\cosh \left( 2\beta _{a}^{\left( \alpha \right) }\tau \right)
-\sinh \left( 2\beta _{a}^{\left( \alpha \right) }\tau \right) +\frac{\left(
B_{a}^{\left( \alpha \right) }\right) ^{2}}{8\left( \beta _{a}^{\left(
\alpha \right) }\right) ^{2}}}\text{,} \\ 
\\ 
\sigma _{a}^{\left( \alpha \right) }\left( \tau \right) =\frac{B_{a}^{\left(
\alpha \right) }\exp \left( -\beta _{a}^{\left( \alpha \right) }\tau \right) 
}{\exp \left( -2\beta _{a}^{\left( \alpha \right) }\tau \right) +\frac{%
\left( B_{a}^{\left( \alpha \right) }\right) ^{2}}{8\left( \beta
_{a}^{\left( \alpha \right) }\right) ^{2}}}+C_{a}^{\left( \alpha \right) }%
\text{.}%
\end{array}%
\end{equation}%
The quantities $B_{a}^{\left( \alpha \right) }$, $C_{a}^{\left( \alpha
\right) }$, $\beta _{a}^{\left( \alpha \right) }$ are real integration
constants and they can be evaluated once the boundary conditions are
specified. We observe that since every geodesic is well-defined for all
temporal parameters $\tau $, $\mathcal{M}_{s}$ constitutes a geodesically
complete manifold \cite{lee}.\textbf{\ }It is therefore a natural setting
within which one may consider global questions and search for a weak
criterion of chaos\textbf{\ }\cite{cipriani}. Furthermore, since $\left\vert
\mu _{a}^{\left( \alpha \right) }\left( \tau \right) \right\vert <+\infty $
and $\left\vert \sigma _{a}^{\left( \alpha \right) }\left( \tau \right)
\right\vert <+\infty $ $\forall \tau \in 
\mathbb{R}
^{+}$, $\forall a=1$, $2$, $3$ and $\forall \alpha =1$,.., $N$, the
parameter space $\left\{ \vec{\Theta}\right\} $ (homeomorphic to $\mathcal{M}%
_{s}$) is compact. The compactness of the configuration space manifold $%
\mathcal{M}_{s}$ assures the folding mechanism of information-dynamical
trajectories (the folding mechanism is a key-feature of true chaos, \cite%
{cipriani}\textbf{).}

It is known \cite{arnold} that the Riemannian curvature of a manifold is
intimately related to the behavior of geodesics on it. If the Riemannian
curvature of a manifold is negative, geodesics (initially parallel) rapidly
diverge from one another. For the sake of simplicity, we assume very special
initial conditions: $B_{a}^{\left( \alpha \right) }\equiv \Xi $, $\beta
_{a}^{\left( \alpha \right) }\equiv \lambda \in 
\mathbb{R}
^{+}$, $C_{a}^{\left( \alpha \right) }=0$, $\forall \alpha =1$, $2$,$....$, $%
l$ and $a=1$, $2$, $3$. However, the conclusions drawn can be generalized to
more arbitrary initial conditions. We observe that since every maximal
geodesic is well-defined for all temporal parameters $\tau $, $\mathcal{M}%
_{s}$ constitute a geodesically complete manifold \cite{lee}. It is
therefore a natural setting within which one may consider global questions
and search for a weak criterion of chaos \cite{cipriani}.

\subsection{Exponential divergence of the Jacobi field intensity}

The actual interest of the Riemannian formulation of the dynamics stems form
the possibility of studying the instability of natural motions through the
instability of geodesics of a suitable manifold, a circumstance that has
several advantages. First of all a powerful mathematical tool exists to
investigate the stability or instability of a geodesic flow: the
Jacobi-Levi-Civita equation for geodesic spread \cite{carmo}. The
JLC-equation describes covariantly how nearby geodesics locally scatter. It
is a familiar object both in Riemannian geometry and theoretical physics (it
is of fundamental interest in experimental General Relativity). Moreover the
JLC-equation relates the stability or instability of a geodesic flow with
curvature properties of the ambient manifold, thus opening a wide and
largely unexplored field of investigation of the connections among geometry,
topology and geodesic instability, hence chaos.

Consider the behavior of the one-parameter family of neighboring geodesics $%
\mathcal{F}_{G_{\mathcal{M}_{s}}}\left( \lambda \right) \equiv \left\{
\Theta _{\mathcal{M}_{s}}^{\mu }\left( \tau \text{; }\lambda \right)
\right\} _{\lambda \in 
\mathbb{R}
^{+}}^{\mu =1\text{,.., }6l}$ where%
\begin{eqnarray}
\mu _{a}^{\left( \alpha \right) }\left( \tau \text{; }\lambda \right) &=&%
\frac{\Xi ^{2}}{2\lambda }\frac{1}{\cosh \left( 2\lambda \tau \right) -\sinh
\left( 2\lambda \tau \right) +\frac{\Xi ^{2}}{8\lambda ^{2}}}\text{,}  \notag
\\
&&  \label{solns} \\
\sigma _{a}^{\left( \alpha \right) }\left( \tau \text{; }\lambda \right)
&=&\Xi \frac{\cosh \left( \lambda \tau \right) -\sinh \left( \lambda \tau
\right) }{\cosh \left( 2\lambda \tau \right) -\sinh \left( 2\lambda \tau
\right) +\frac{\Xi ^{2}}{8\lambda ^{2}}}\text{.}  \notag
\end{eqnarray}%
with $\alpha =1$, $2$,$....$, $l$ and $a=1$, $2$, $3$. The relative geodesic
spread on a (non-maximally symmetric) curved manifold as $\mathcal{M}_{s}$
is characterized by the Jacobi-Levi-Civita equation, the natural tool to
tackle dynamical chaos \cite{clarke, carmo},%
\begin{equation}
\frac{D^{2}\delta \Theta ^{\mu }}{D\tau ^{2}}+R_{\nu \rho \sigma }^{\mu }%
\frac{\partial \Theta ^{\nu }}{\partial \tau }\delta \Theta ^{\rho }\frac{%
\partial \Theta ^{\sigma }}{\partial \tau }=0  \label{gen-geoDev}
\end{equation}%
where the covariant derivative $\frac{D^{2}\delta \Theta ^{\mu }}{D\tau ^{2}}
$\textbf{\ }in (\ref{gen-geoDev}) is defined as \cite{ohanian},%
\begin{eqnarray}
\frac{D^{2}\delta \Theta ^{\mu }}{D\tau ^{2}} &=&\frac{d^{2}\delta \Theta
^{\mu }}{d\tau ^{2}}+2\Gamma _{\alpha \beta }^{\mu }\frac{d\delta \Theta
^{\alpha }}{d\tau }\frac{d\Theta ^{\beta }}{d\tau }+\Gamma _{\alpha \beta
}^{\mu }\delta \Theta ^{\alpha }\frac{d^{2}\Theta ^{\beta }}{d\tau ^{2}}%
+\Gamma _{\alpha \beta ,\nu }^{\mu }\frac{d\Theta ^{\nu }}{d\tau }\frac{%
d\Theta ^{\beta }}{d\tau }\delta \Theta ^{\alpha }+  \notag \\
&&+\Gamma _{\alpha \beta }^{\mu }\Gamma _{\rho \sigma }^{\alpha }\frac{%
d\Theta ^{\sigma }}{d\tau }\frac{d\Theta ^{\beta }}{d\tau }\delta \Theta
^{\rho }\text{,}
\end{eqnarray}%
and the Jacobi vector field $J^{\mu }$ is given by \cite{defelice},%
\begin{equation}
J^{\mu }\equiv \delta \Theta ^{\mu }\overset{\text{def}}{=}\delta _{\lambda
}\Theta ^{\mu }=\left. \left( \frac{\partial \Theta ^{\mu }\left( \tau \text{%
; }\lambda \right) }{\partial \lambda }\right) \right\vert _{\tau =\text{%
const}}\delta \lambda \text{.}  \label{jacobi}
\end{equation}%
Notice that the JLC-equation appears intractable already at rather small $l$%
. For isotropic manifolds, the JLC-equation can be reduced to the simple
form \cite{carmo},%
\begin{equation}
\frac{D^{2}J^{\mu }}{D\tau ^{2}}+KJ^{\mu }=0\text{, }\mu =1\text{,...., }6l
\label{geo-deviation}
\end{equation}%
where $K$ is the constant value assumed throughout the manifold by the
sectional curvature. The sectional curvature of manifold $\mathcal{M}_{s}$
is the $6l$-dimensional generalization of the Gaussian curvature of
two-dimensional surfaces of $%
\mathbb{R}
^{3}$. If $K<0$, unstable solutions of the equation (\ref{geo-deviation})
are of the form%
\begin{equation}
J\left( \tau \right) =\frac{1}{\sqrt{-K}}\omega \left( 0\right) \sinh \left( 
\sqrt{-K}\tau \right)
\end{equation}%
once the initial conditions are assigned as $J\left( 0\right) =0$, $\frac{%
dJ\left( 0\right) }{d\tau }=\omega \left( 0\right) $ and $K<0$. Equation (%
\ref{gen-geoDev}) forms a system of $6l$ coupled ordinary differential
equations linear in the components of the deviation vector field (\ref%
{jacobi}) but\textit{\ }nonlinear in derivatives of the metric (\ref%
{fisher-rao}). It describes the linearized geodesic flow: the linearization
ignores the relative velocity of the geodesics. When the geodesics are
neighboring but their relative velocity is arbitrary, the corresponding
nonlinear geodesic deviation equation is the so-called generalized Jacobi
equation \cite{chicone, hodgkinson}. The nonlinearity is due to the
existence of velocity-dependent terms in the system. Neighboring geodesics
accelerate relative to each other with a rate directly measured by the
curvature tensor $R_{\alpha \beta \gamma \delta }$. Substituting (\ref{solns}%
) in (\ref{gen-geoDev}) and neglecting the exponentially decaying terms in $%
\delta \Theta ^{\mu }$ and its derivatives, integration of (\ref{gen-geoDev}%
) leads to the following asymptotic exponential growth of the Jacobi vector
field intensity (a classical feature of chaos),%
\begin{equation}
J_{\mathcal{M}_{S}}=\left\Vert J\right\Vert =\left( g_{\mu \nu }J^{\mu
}J^{\nu }\right) ^{\frac{1}{2}}\overset{\tau \rightarrow \infty }{\approx }%
3le^{\lambda \tau }\text{.}
\end{equation}%
Finally, we point out that in our approach the quantity $\lambda _{J}$,%
\begin{equation}
\lambda _{J}\overset{\text{def}}{=}\underset{\tau \rightarrow \infty }{\lim }%
\frac{1}{\tau }\ln \left[ \frac{\left\Vert J_{_{\mathcal{M}_{S}}}\left( \tau
\right) \right\Vert }{\left\Vert J_{_{\mathcal{M}_{S}}}\left( 0\right)
\right\Vert }\right]
\end{equation}%
would play the role of the conventional Lyapunov exponents.

\section{Linearity of the information geometrodynamical entropy}

We investigate the stability of the trajectories of the ED model considered
on $\mathcal{M}_{s}$. It is known \cite{arnold} that the Riemannian
curvature of a manifold is closely connected with the behavior of the
geodesics on it. If the Riemannian curvature of a manifold is negative,
geodesics (initially parallel) rapidly diverge from one another. For the
sake of simplicity, we assume very special initial conditions: $%
B_{a}^{\left( \alpha \right) }\equiv \Xi $, $\beta _{a}^{\left( \alpha
\right) }\equiv \lambda \in 
\mathbb{R}
^{+}$, $C_{a}^{\left( \alpha \right) }=0$, $\forall $ $\alpha =1$, $2$,$....$%
, $l$ and $a=1$, $2$, $3$ . However, the conclusion we reach can be
generalized to more arbitrary initial conditions. Recall $\mathcal{M}_{s}$
is the space of probability distributions $\left\{ P\left( \vec{X}\left\vert 
\vec{\Theta}\right. \right) \right\} $ labeled by $6l$ statistical
parameters $\vec{\Theta}$. These parameters are the coordinates for the
point $P$, and in these coordinates a volume element $dV_{\mathcal{M}_{s}}$
reads, 
\begin{equation}
dV_{\mathcal{M}_{S}}=\sqrt{g}d^{6l}\vec{\Theta}=\dprod\limits_{\alpha
=1}^{l}\dprod\limits_{a=1}^{3}\frac{\sqrt{2}}{\left( \sigma _{a}^{\left(
\alpha \right) }\right) ^{2}}d\mu _{a}^{\left( \alpha \right) }d\sigma
_{a}^{\left( \alpha \right) }\text{.}
\end{equation}%
The volume of an extended region $\Delta V_{\mathcal{M}_{s}}\left( \tau 
\text{; }\lambda \right) $ of $\mathcal{M}_{s}$ is defined by,%
\begin{equation}
\Delta V_{\mathcal{M}_{s}}\left( \tau \text{; }\lambda \right) \overset{%
\text{def}}{=}\dprod\limits_{\alpha
=1}^{l}\dprod\limits_{a=1}^{3}\int\nolimits_{\mu _{a}^{\left( \alpha \right)
}\left( 0\right) }^{\mu _{a}^{\left( \alpha \right) }\left( \tau \right)
}\int\nolimits_{\sigma _{a}^{\left( \alpha \right) }\left( 0\right)
}^{\sigma _{a}^{\left( \alpha \right) }\left( \tau \right) }\frac{\sqrt{2}}{%
\left( \sigma _{a}^{\left( \alpha \right) }\right) ^{2}}d\mu _{a}^{\left(
\alpha \right) }d\sigma _{a}^{\left( \alpha \right) }
\end{equation}%
where $\mu _{a}^{\left( \alpha \right) }\left( \tau \right) $ and $\sigma
_{a}^{\left( \alpha \right) }\left( \tau \right) $ are given in (\ref{solns}%
) and where the scalar $\lambda $ is the chosen quantity used to define the
one-parameter family of geodesics $\mathcal{F}_{G_{\mathcal{M}_{s}}}\left(
\lambda \right) \overset{\text{def}}{=}\left\{ \Theta _{\mathcal{M}%
_{s}}^{\mu }\left( \tau \text{; }\lambda \right) \right\} _{\lambda \in 
\mathbb{R}
^{+}}^{\mu =1\text{,..,}6l}$. The quantity that encodes relevant information
about the stability of neighboring volume elements is the the average volume 
$\mathcal{V}_{\mathcal{M}_{s}}\left( \tau \text{; }\lambda \right) $, 
\begin{equation}
\mathcal{V}_{\mathcal{M}_{s}}\left( \tau \text{; }\lambda \right) \equiv
\left\langle \Delta V_{\mathcal{M}_{s}}\left( \tau ^{\prime }\text{; }%
\lambda \right) \right\rangle _{\tau }\overset{\text{def}}{=}\frac{1}{\tau }%
\dint\limits_{0}^{\tau }\Delta V_{\mathcal{M}_{s}}\left( \tau ^{\prime }%
\text{; }\lambda \right) d\tau ^{\prime }\overset{\tau \rightarrow \infty }{%
\approx }e^{3l\lambda \tau }\text{.}  \label{avg-vol}
\end{equation}%
The ratio $\frac{\mathcal{V}_{\mathcal{M}_{s}}\left( \tau \right) }{\mathcal{%
V}_{\mathcal{M}_{s}}\left( 0\right) }$ with $\mathcal{V}_{\mathcal{M}%
_{s}}\left( \tau \right) $\textbf{\ }in (\ref{avg-vol}) representing the
temporal average of the\textbf{\ }$3l$\textbf{-}fold integral over
trajectories of maximum probability (geodesics) is a measure of the number
of the accessible macrostates in configuration (statistical) manifold $%
\mathcal{M}_{s}$ after a finite temporal increment\textbf{\ }$\tau $. In
other words, $\mathcal{V}_{\mathcal{M}_{s}}\left( \tau \right) $ can be
interpreted as the temporal evolution of the system's uncertainty volume $%
\mathcal{V}_{\mathcal{M}_{s}}\left( 0\right) $. For instance $\mathcal{V}_{%
\mathcal{M}_{s}}\left( 0\right) $ may be a spherical volume of initial
points whose center is a given point on the attractor and whose surface
consists of configuration points from nearby trajectories. An attractor is a
subset of the manifold $\mathcal{M}_{s}$\textbf{\ }toward which almost all
sufficiently close trajectories converge asymptotically, covering it densely
as the time goes on. Strange attractors are called chaotic attractors.
Chaotic attractors have at least one finite positive Lyapunov exponent \cite%
{tel}. As the center of $\mathcal{V}_{\mathcal{M}_{s}}\left( 0\right) $ and
its surface points evolve in time, the spherical volume becomes an ellipsoid
with principal axes in the directions of contraction and expansion. The
average rates of expansion and contraction along the principal axes are the
Lyapunov exponents \cite{wolfA}.

The asymptotic regime of diffusive evolution in (\ref{avg-vol}) describes
the exponential increase of average volume elements on $\mathcal{M}_{s}$.
The exponential instability characteristic of chaos forces the system to
rapidly explore large areas (volumes) of the statistical manifolds. From
equation (\ref{avg-vol}), we notice that the parameter $\lambda $
characterizes the exponential growth rate of average statistical volumes $%
\mathcal{V}_{\mathcal{M}_{s}}\left( \tau \text{; }\lambda \right) $ in $%
\mathcal{M}_{s}$. This suggests that $\lambda $ may play the same role
ordinarily played by Lyapunov exponents \cite{ruelle}. It is interesting to
note that this asymptotic behavior appears also in the conventional
description of quantum chaos where the von Neumann entropy increases
linearly at a rate determined by the Lyapunov exponents. The linear increase
of entropy as a quantum chaos criterion was introduced by Zurek and Paz \cite%
{zurek}. In our information-geometric approach a relevant quantity that can
be useful to study the degree of instability characterizing the ED model is
the information-geometrodynamical entropy (IGE) defined as \cite{cafaro2},%
\begin{equation}
\mathcal{S}_{\mathcal{M}_{s}}\overset{\text{def}}{=}\underset{\tau
\rightarrow \infty }{\lim }\log \mathcal{V}_{\mathcal{M}_{s}}\left( \tau 
\text{; }\lambda \right) \text{.}  \label{asym-ent}
\end{equation}%
The IGE\ is intended to capture the temporal complexity (chaoticity) of ED\
theoretical models on curved statistical manifolds $\mathcal{M}_{s}$ by
considering the asymptotic temporal behaviors of the average statistical
volumes occupied by the evolving macrovariables labelling points on $%
\mathcal{M}_{s}$. Substituting (\ref{avg-vol}) in (\ref{asym-ent}), we obtain%
\begin{equation}
\mathcal{S}_{\mathcal{M}_{s}}=\underset{\tau \rightarrow \infty }{\lim }\log
\left\{ \frac{1}{\tau }\dint\limits_{0}^{\tau }\left[ \dprod\limits_{\alpha
=1}^{l}\dprod\limits_{a=1}^{3}\int\nolimits_{\mu _{a}^{\left( \alpha \right)
}\left( 0\right) }^{\mu _{a}^{\left( \alpha \right) }\left( \tau ^{\prime
}\right) }\int\nolimits_{\sigma _{a}^{\left( \alpha \right) }\left( 0\right)
}^{\sigma _{a}^{\left( \alpha \right) }\left( \tau ^{\prime }\right) }\frac{%
\sqrt{2}}{\left( \sigma _{a}^{\left( \alpha \right) }\right) ^{2}}d\mu
_{a}^{\left( \alpha \right) }d\sigma _{a}^{\left( \alpha \right) }\right]
d\tau ^{\prime }\right\} \overset{\tau \rightarrow \infty }{\approx }%
3l\lambda \tau \text{.}  \label{ent-Ms}
\end{equation}%
Before discussing the meaning of (\ref{ent-Ms}), recall that in conventional
approaches to chaos the notion of entropy is introduced, in both classical
and quantum physics, as the missing information about the systems
fine-grained state \cite{caves, jaynes}. For a classical system, suppose
that the phase space is partitioned into very fine-grained cells of uniform
volume $\Delta v$, labelled by an index $j$. If one does not know which cell
the system occupies, one assigns probabilities $p_{j}$ to the various cells;
equivalently, in the limit of infinitesimal cells, one can use a phase-space
density $\rho \left( X_{j}\right) =\frac{p_{j}}{\Delta v}$. Then, in a
classical chaotic evolution, the asymptotic expression of the information
needed to characterize a particular coarse-grained trajectory out to time $%
\tau $ is given by the Shannon information entropy (measured in bits),%
\begin{equation}
\mathcal{S}_{\text{classical}}^{\left( \text{chaotic}\right) }=-\int dX\rho
\left( X\right) \log _{2}\left( \rho \left( X\right) \Delta v\right)
=-\sum_{j}p_{j}\log _{2}p_{j}\sim \mathcal{K}\tau \text{.}
\label{chao-classEnt}
\end{equation}%
where $\rho \left( X\right) $ is the phase-space density and $p_{j}=\frac{%
v_{j}}{\Delta v}$ is the probability for the corresponding coarse-grained
trajectory. $\mathcal{S}_{\text{classical}}^{\left( \text{chaotic}\right) }$
is the missing information about which fine-grained cell the system
occupies. The quantity $\mathcal{K}$ represents the linear rate of
information increase and it is called the Kolmogorov-Sinai entropy (or
metric entropy) ($\mathcal{K}$ is the sum of positive Lyapunov exponents, $%
\mathcal{K}=\sum_{j}\lambda _{j}$ ). $\mathcal{K}$ quantifies the degree of
classical chaos. The Kolmogorov-Sinai entropy provides a measure of the rate
at which information is lost by an evolving chaotic system ( $\mathcal{K}$
has dimension entropy/time) and has its roots in the definition of the
Shannon entropy.\textbf{\ }It is worthwhile emphasizing that the quantity
that grows asymptotically as $\mathcal{K}\tau $ is really the average of the
information on the left side of equation (\ref{chao-classEnt}). This
distinction can however be ignored provided we assume the chaotic system has
roughly constant Lyapunov exponents over the accessible region of phase
space. In quantum mechanics the fine-grained alternatives are normalized
state vectors in Hilbert space. From a set of probabilities for various
state vectors, one can construct a density operator 
\begin{equation}
\widehat{\rho }=\sum_{j}\lambda _{j}\left\vert \psi _{j}\right\rangle
\left\langle \psi _{j}\right\vert \text{, }\widehat{\rho }\left\vert \psi
_{j}\right\rangle =\lambda _{j}\left\vert \psi _{j}\right\rangle \text{.}
\end{equation}%
The normalization of the density operator, $tr\left( \widehat{\rho }\right)
=1$, implies that the eigenvalues make up a normalized probability
distribution. The von Neumann entropy (natural generalization of both
Boltzmann's and Shannon's entropy) of the density operator $\widehat{\rho }$
(measured in bits) \cite{stenholm},%
\begin{equation}
\mathcal{S}_{\text{quantum}}^{\left( \text{chaotic}\right) }=-tr\left( 
\widehat{\rho }\log _{2}\widehat{\rho }\right) =-\sum_{j}\lambda _{j}\log
_{2}\lambda _{j}\sim \mathcal{K}_{q}\tau  \label{squantum}
\end{equation}%
can be thought of as the missing information about which eigenvector the
system is in. Entropy quantifies the degree of unpredictability about the
system's fine-grained state. In quantum mechanics, the von Neumann entropy
plays a role analogous to that played by the Shannon entropy in classical
probability theory. They are both functionals of the state, are both
monotone under a relevant kind of mapping, and can both be singled out
uniquely by natural requirements. von Neumann's entropy reduces to the
Shannon entropy for diagonal density matrices. However, in general the von
Neumann entropy is a subtler object than its classical counterpart. The
quantity $\mathcal{K}_{q}$\ in (\ref{squantum}) can be interpreted as the
non-commutative (quantum theory is a non-commutative probability theory)
quantum analog of the Kolmogorov-Sinai dynamical entropy, the so-called
quantum dynamical entropy \cite{benatti}. Examples of quantum dynamical
entropies applied to quantum chaos and quantum information theory are the
Alicki-Fannes (AF) \cite{alicki} entropy and the Connes-Narnhofer-Thirring
(CNT) \cite{connes} entropy. Both the AF and CNT entropy coincide with the
KS entropy on classical dynamical systems. They also coincide on
finite-dimensional quantum systems. However, they differ when moving from
finite to infinite quantum systems.

Recall that decoherence is the loss of phase coherence between the set of
preferred quantum states in the Hilbert space of the system due to the
interaction with the environment. Moreover, decoherence induces transitions
from quantum to classical systems. Therefore, classicality is an emergent
property of an open quantum system. Motivated by such considerations, Zurek
and Paz investigated implications of the process of decoherence for quantum
chaos. They considered a chaotic system, a single unstable harmonic
oscillator characterized by a potential $V\left( x\right) =-\frac{\lambda
x^{2}}{2}$ ($\lambda $ is the Lyapunov exponent), coupled to an external
environment. In the reversible classical limit\textit{\ }\cite{zurek2}, the
von Neumann entropy of such a system increases linearly at a rate determined
by the Lyapunov exponent,%
\begin{equation}
\mathcal{S}_{\text{quantum}}^{\left( \text{chaotic}\right) }\left( \text{%
Zurek-Paz}\right) \overset{\tau \rightarrow \infty }{\sim }\lambda \tau 
\text{.}
\end{equation}%
Notice that the consideration of $3l$ uncoupled identical unstable harmonic
oscillators characterized by potentials $V_{i}\left( x\right) =-\frac{%
\lambda _{i}x^{2}}{2}$ $\left( \lambda _{i}=\lambda _{j}\text{; }i\text{, }%
j=1\text{, }2\text{,..., }3l\right) $ would simply lead to%
\begin{equation}
\mathcal{S}_{\text{quantum}}^{\left( \text{chaotic}\right) }\left( \text{%
Zurek-Paz}\right) \overset{\tau \rightarrow \infty }{\sim }3l\lambda \tau 
\text{.}  \label{other-ent}
\end{equation}%
The resemblance of equations (\ref{ent-Ms}) and (\ref{other-ent}) is
remarkable and a more detailed discussion about this analogy is presented in 
\cite{cafaro7} where an information-geometric analogue of the Zurek-Paz
quantum chaos criterion in the classical reversible limit is proposed \cite%
{zurek}. This analogy is illustrated applying the IGAC to a set of $n$%
-uncoupled three-dimensional anisotropic inverted harmonic oscillators
characterized by a Ohmic distributed frequency spectrum.

The entropy-like quantity $\mathcal{S}_{\mathcal{M}_{s}}$ in (\ref{ent-Ms})
is the asymptotic limit of the natural logarithm of the statistical weight $%
\left\langle \Delta V_{\mathcal{M}_{s}}\right\rangle _{\tau }$ defined on $%
\mathcal{M}_{s}$ and it grows linearly in time, a quantum feature of chaos.
\ Indeed, equation (\ref{ent-Ms}) may be considered the
information-geometric analog of the Zurek-Paz chaos criterion. In our
chaotic ED\ Gaussian model, the IGE production rate is determined by the
information-geometric parameter $\lambda $ characterizing the exponential
growth rate of average statistical volumes $\mathcal{V}_{\mathcal{M}%
_{s}}\left( \tau \text{; }\lambda \right) $ in $\mathcal{M}_{s}$.

In conclusion, for the example under investigation, we have%
\begin{equation}
\mathcal{R}_{\mathcal{M}_{s}}=-3l\text{, }\mathcal{S}_{\mathcal{M}_{s}}%
\overset{\tau \rightarrow \infty }{\approx }3l\lambda \tau \text{, }J_{%
\mathcal{M}_{S}}\overset{\tau \rightarrow \infty }{\approx }3le^{\lambda
\tau }\text{.}  \label{ee}
\end{equation}%
The IGE grows linearly as a function of the number of Gaussian-distributed
microstates of the system. This supports the fact that $\mathcal{S}_{%
\mathcal{M}_{s}}$\ may be a useful measure of temporal complexity \cite{JP}.
Furthermore, these three indicators of chaoticity, the Ricci scalar
curvature $\mathcal{R}_{\mathcal{M}_{s}}$, the information-geometric entropy 
$\mathcal{S}_{\mathcal{M}_{s}}$ and the Jacobi vector field intensity $J_{%
\mathcal{M}_{S}}$ are proportional to $3l$, the dimension of the microspace
with microstates $\left\{ \vec{X}\right\} $ underlying our chaotic ED\
Gaussian model. This proportionality leads to the conclusion that there is a
substantial link among these information-geometric measures of chaoticity
since they are all extensive functions of the dimensionality of the
microspace underlying the macroscopic chaotic entropic dynamics (see (\ref%
{ee})). Curvature, information-geometrodynamical entropy and Jacobi field
intensity are linked within our formalism. We are aware that our findings
are reliable in the restrictive assumption of Gaussianity. However, we
believe that with some additional technical machinery, more general
conclusions can be achieved and this connection among indicators of
chaoticity may be strengthened.

\section{Information geometry of quantum energy level statistics: An
application to entanglement in quantum spin chains}

In what follows, we apply the IGAC to study the entropic dynamics on curved
statistical manifolds induced by classical probability distributions of
common use in the study of regular and chaotic quantum energy level
statistics. In doing so, we suggest an information-geometric
characterization of a special class of regular and chaotic quantum energy
level statistics. More precisely, we present an information-geometric
analogue of the logarithmic and linear entanglement entropy growth in
regular and quantum chaotic spin chains, respectively.

\subsection{The Information Geometry of the Poisson and Wigner-Dyson
Distributions}

The theory of quantum chaos (quantum mechanics of systems whose classical
dynamics are chaotic) is not primarily related to few-body physics. Indeed,
in real physical systems such as many-electron atoms and heavy nuclei, the
origin of complex behavior is the quite strong interaction among many
particles. To deal with such systems, a famous statistical approach has been
developed which is based upon the Random Matrix Theory (RMT) \cite{porter}.
The main idea of this approach is to neglect the detailed description of the
motion and to treat these systems statistically bearing in mind that the
interaction among particles is so complex and strong that generic properties
are expected to emerge. The simplest models of RMT\ are full random matrices
of a given symmetry. One of the main results of RMT is the prediction of a
specific kind of correlations of the energy spectra of complex quantum
systems. Among many characteristics of these correlations, the most popular
one is the distribution of spacings between nearest energy levels in the
spectra. The exact analytical expression of this distribution is very
complicated; instead, one uses the so-called Wigner-Dyson surmise (a very
simple expression which gives a very good approximation to the exact
result). The known manifestation of quantum chaos is the so-called
Wigner-Dyson (WD) distribution for spacings between neighboring levels in
the spectrum. In the other limiting case of completely integrable (regular)
systems, the distribution turns out to be very close to the Poissonian one.
A distinctive property of the WD distribution is the repulsion between
neighboring levels in the spectra; the degree of this repulsion (linear,
quadratic or quartic) depends on the symmetry of random matrices. For
systems without time reversal invariance the relevant ensemble of random
matrices is the Gaussian Unitary Ensemble (GUE) \cite{porter}, characterized
by the probability distribution%
\begin{equation}
p_{\text{GUE}}\left( \theta \right) =\frac{32}{\pi ^{2}}\theta ^{2}\exp
\left( -\frac{4}{\pi }\theta ^{2}\right) \text{, (quadratic repulsion)}
\label{GUE}
\end{equation}%
where $\theta $ is the average spacing of the energy levels.\textbf{\ }For
systems invariant with respect to time reversal the ensemble is the Gaussian
Orthogonal Ensemble (GOE) \cite{porter},%
\begin{equation}
p_{\text{GOE}}\left( \theta \right) =\frac{\pi }{2}\theta \exp \left( -\frac{%
\pi }{4}\theta ^{2}\right) \text{, (linear repulsion).}  \label{GOE}
\end{equation}%
For systems with time reversal invariance but with half-integer spin, the
energy is described by the Gaussian Symplectic Ensemble (GSE) of random
matrices \cite{porter},%
\begin{equation}
p_{\text{GSE}}\left( \theta \right) =\frac{2^{18}}{3^{6}\pi ^{3}}\theta
^{4}\exp \left( -\frac{64}{9\pi }\theta ^{2}\right) \text{, (quartic
repulsion).}  \label{GSE}
\end{equation}%
Equations (\ref{GUE}), (\ref{GOE}) and (\ref{GSE}) are standard accepted
conjectures. Besides energy level statistics in the extreme integrable
(Poisson) and chaotic (Wigner-Dyson) regimes, there is also energy level
statistics in the mixed regime, i.e., such having a mixed classical dynamics
where regular and chaotic regions coexist in the phase space. A convenient
and often successful parametrization of the correct probability distribution
in the transition region between Poisson and WD distributions is provided by
the Brody interpolation formula \cite{brody},%
\begin{equation}
p_{\beta }^{\left( \text{Brody}\right) }\left( \theta \right) =\gamma \left(
\beta +1\right) \exp \left( -\gamma \theta ^{\beta +1}\right) \text{.}
\end{equation}%
where $\gamma =\left\{ \Gamma \left[ \frac{\beta +2}{\beta +1}\right]
\right\} ^{\beta +1}$\ and $\Gamma \left( \beta \right) $\ is the Euler
Gamma function. This distribution is normalized and, by construction, has
mean spacing $\left\langle \theta \right\rangle =1$. We recover the Poisson
case by taking $\beta =0$\ while the Wigner case is recovered for $\beta =1$%
. However, a criticism of the Brody distribution is the lack of a first
principles justification for its validity. The fact remains that it does fit
the specific results found when considering explicit model systems. It is
essentially an ad hoc one-parameter family of distributions and has no deep
physical background, but it does interpolate between Poisson and
Wigner-Dyson in a simple, effective manner. Our objective here is to apply
our information-geometric formalism (based on statistical inference methods)
to Wigner-Dyson and Poisson probability distributions.

Most of the probability distributions arise from the maximum entropy
formalism as a result of some simple statements concerning averages. Not all
distribution are generated in this way. Some distributions are generated by
combining the results of simple cases (multinomial from a binomial). Other
distributions are found as a result of a change of variable (Cauchy
distribution). For instance, the Weibull distribution \cite{tribus} can be
obtained from an exponential distribution as a result of a power law
transformation. Assume our knowledge of the microstate $x$ is encoded in an
exponential distribution,%
\begin{equation}
p\left( x|\theta \right) =\frac{1}{\theta }e^{-\frac{x}{\theta }}\text{,}
\label{uno}
\end{equation}%
where $x$ may be considered the spacing of the energy levels while $\theta $
is the average spacing, $\theta =\left\langle x\right\rangle $. Note that
the study of probability distributions could, in principle, be restricted to
the exponential type since an arbitrary distribution can be represented in
exponential form. It is said that the exponential family of distributions is
dense in the totality of probability distributions \cite{brody-exp}.\textbf{%
\ }We can re-express $x\in \mathcal{X}$ in $p\left( x|\theta \right) $ in
terms of another random variable $y=f\left( x\right) \in \mathcal{Y}$,
assuming $f$ is an invertible mapping. For instance, consider the power law
transformation%
\begin{equation}
x\rightarrow y=f\left( x\right) =\left( \frac{x}{\zeta }\right) ^{\frac{1}{n}%
}\text{.}  \label{due}
\end{equation}%
We clearly have,%
\begin{eqnarray}
p_{\text{old}}\left( x\right) &\rightarrow &\hat{p}_{\text{new}}\left(
y\right) =\underset{\mathcal{X}}{\int }dxp_{\text{old}}\left( x\right)
\delta \left( y-f\left( x\right) \right)  \notag \\
&&  \notag \\
&=&\underset{\mathcal{X}}{\int }dxp_{\text{old}}\left( x\right) \frac{1}{%
\left\vert \frac{\partial f}{\partial x}\right\vert }\delta \left(
f^{-1}\left( y\right) -x\right) =\left[ \frac{1}{\left\vert \frac{\partial f%
}{\partial x}\right\vert }p_{\text{old}}\left( x\right) \right]
_{x=f^{-1}\left( y\right) }\text{.}  \label{tre}
\end{eqnarray}%
Therefore, considering (\ref{uno}) and (\ref{due}), equation (\ref{tre})
leads to%
\begin{equation}
\hat{p}_{\text{new}}\left( y\right) =n\frac{\zeta }{\theta }e^{-\frac{\zeta 
}{\theta }y^{n}}y^{n-1}\text{.}  \label{new}
\end{equation}%
It is worthwhile emphasizing that since $\left\vert \frac{\partial f}{%
\partial x}\right\vert $ does not depend on $\theta $ and since $\underset{%
\mathcal{Y}}{\int }dy=\underset{\mathcal{X}}{\int }dx\left\vert \frac{%
\partial f}{\partial x}\right\vert $, we have%
\begin{equation}
\underset{\mathcal{Y}}{\int }dy\hat{p}_{\text{new}}\left( y\right) \partial
_{\mu }\log \hat{p}_{\text{new}}\left( y\right) \partial _{\nu }\log \hat{p}%
_{\text{new}}\left( y\right) =\underset{\mathcal{X}}{\int }dyp_{\text{old}%
}\left( x\right) \partial _{\mu }\log p_{\text{old}}\left( x\right) \partial
_{\nu }\log p_{\text{old}}\left( x\right) \text{.}  \label{invariance}
\end{equation}%
Equation (\ref{invariance}) leads to conclude that the Fisher-Rao
information metric $g_{\mu \nu }$ is invariant under transformations of the
random variable. For the sake of completeness, let us show that the
information metric is also covariant under reparametrization. Suppose that $%
\left( \hat{\theta}_{\mu }\right) $ is a new set of coordinates, specified
in terms of the old set through the invertible relationship $\hat{\theta}%
_{\mu }=\hat{\theta}_{\mu }\left( \theta \right) $. Defining $\hat{p}_{_{%
\hat{\theta}}}\left( x\right) \equiv p_{_{\theta \left( \hat{\theta}\right)
}}\left( x\right) $, we are then able to compute the new metric tensor $\hat{%
g}_{\mu \nu }\left( \hat{\theta}\right) $ in terms of $g_{\mu \nu }\left(
\theta \right) $. Indeed, since $\frac{\partial }{\partial \hat{\theta}^{\mu
}}\hat{p}_{_{\hat{\theta}}}=\frac{\partial \theta ^{\nu }}{\partial \hat{%
\theta}^{\mu }}$ $\frac{\partial }{\partial \theta ^{\nu }}p_{_{\theta
\left( \hat{\theta}\right) }}$, we obtain%
\begin{equation}
\hat{g}_{\mu \nu }\left( \hat{\theta}\right) =\left[ \frac{\partial \theta
^{\rho }}{\partial \hat{\theta}^{\mu }}\frac{\partial \theta ^{\sigma }}{%
\partial \hat{\theta}^{\nu }}g_{\rho \sigma }\left( \theta \right) \right]
_{\theta =\theta \left( \hat{\theta}\right) }\text{.}  \label{covariance}
\end{equation}%
Letting $\frac{\zeta }{\theta }=\frac{1}{\Lambda ^{n}}$, from (\ref{new}) we
obtain the Weibull probability distribution, 
\begin{equation}
p_{\text{Weibull}}\left( y|\Lambda \right) =\frac{n}{\Lambda }\left( \frac{y%
}{\Lambda }\right) ^{n-1}e^{-\left( \frac{y}{\Lambda }\right) ^{n}}\text{, }%
\Lambda =\left( \frac{\theta }{\zeta }\right) ^{\frac{1}{n}}\text{.}
\label{weibull}
\end{equation}%
Moreover, letting $n=2$, $y=\Delta $ and $\Lambda =\frac{2D}{\sqrt{\pi }}$,
from (\ref{weibull}) we obtain the standard Wigner-Dyson distribution,%
\begin{equation}
p_{\text{Wigner-Dyson}}\left( \Delta |D\right) =\frac{\pi \Delta }{2D^{2}}%
e^{-\frac{\pi \Delta ^{2}}{4D^{2}}}\text{, }D=\frac{\sqrt{\pi }}{2}\left( 
\frac{\theta }{\zeta }\right) ^{\frac{1}{2}}\text{.}  \label{uno_1}
\end{equation}%
In conventional notations, $\Delta $ is the spacing between two neighboring
energy levels and $D$ is the average spacing \cite{biro}. Recall that the
Fisher-Rao information metric $G_{\mu \nu }^{\left( \text{P}\right) }\left(
\theta \right) $ of a Poissonian probability distribution $p\left( x|\theta
\right) $ is defined as,%
\begin{equation}
G_{\mu \nu }^{\left( \text{P}\right) }\left( \theta \right) =\int dxp\left(
x|\theta \right) \partial _{\mu }\log p\left( x|\theta \right) \partial
_{\nu }\log p\left( x|\theta \right) \text{ with }\partial _{\mu }=\frac{%
\partial }{\partial \theta ^{\mu }}
\end{equation}%
where $p\left( x|\theta \right) $ is given by,%
\begin{equation}
p\left( x|\theta \right) =\frac{1}{\theta }\exp \left( -\frac{x}{\theta }%
\right) \text{.}
\end{equation}%
The Poisson line element $\left( ds^{2}\right) _{\text{Poisson}}$ is defined
as,%
\begin{equation}
\left( ds^{2}\right) _{\text{Poisson}}=G_{\mu \nu }^{\left( \text{P}\right)
}\left( \theta \right) d\theta ^{\mu }d\theta ^{\nu }=\frac{1}{\theta ^{2}}%
d\theta ^{2}\text{.}  \label{P_line}
\end{equation}%
The Fisher-Rao information metric $G_{\mu \nu }^{(\text{WD})}\left( \phi
\right) $ of a Wigner-Dyson probability distribution $q\left( y|\phi \right) 
$ is defined as,%
\begin{equation}
G_{\mu \nu }^{(\text{WD})}\left( \phi \right) =\int dyq\left( y|\phi \right)
\partial _{\mu }\log q\left( y|\phi \right) \partial _{\nu }\log q\left(
y|\phi \right) \text{ with }\partial _{\mu }=\frac{\partial }{\partial \phi
^{\mu }}\text{ }
\end{equation}%
where $q\left( y|\phi \right) $ is given by,%
\begin{equation}
q\left( y|\phi \right) =\frac{\pi y}{2\phi ^{2}}\exp \left( -\frac{\pi y^{2}%
}{4\phi ^{2}}\right) \text{, }\phi =\frac{\sqrt{\pi }}{2}\left( \frac{\theta 
}{\lambda }\right) ^{\frac{1}{2}}\text{.}  \label{due_2}
\end{equation}%
Notice that $q\left( y|\phi \right) $ in (\ref{due_2}) is equivalent to $p_{%
\text{Wigner-Dyson}}\left( \Delta |D\right) $ in (\ref{uno_1}) with $%
y=\Delta $ and $\phi =D$. The Wigner-Dyson line element $\left(
ds^{2}\right) _{\text{Wigner-Dyson}}$ is defined as,%
\begin{equation}
\left( ds^{2}\right) _{\text{Wigner-Dyson}}=G_{\mu \nu }^{\left( \text{WD}%
\right) }\left( \phi \right) d\phi ^{\mu }d\phi ^{\nu }\text{.}
\end{equation}%
Notice that the Poisson distribution and the Wigner-Dyson distributions are
related through the combination of a change of random variable and a new
reparametrization, namely%
\begin{equation}
q\left( y|\phi \right) =p\left( x\left( y\right) |\theta \left( \phi \right)
\right) J\left( y\right)
\end{equation}%
where,%
\begin{equation}
x\left( y\right) =\lambda y^{2}\text{, }\theta \left( \phi \right) =\frac{%
4\phi ^{2}}{\pi }\lambda \text{, }J\left( y\right) =\left\vert \frac{%
\partial x\left( y\right) }{\partial y}\right\vert \text{.}
\end{equation}%
Considering equations (\ref{invariance}) and (\ref{covariance}), the
Wigner-Dyson line element $\left( ds^{2}\right) _{\text{Wigner-Dyson}}$
becomes%
\begin{eqnarray}
\left( ds^{2}\right) _{\text{Wigner-Dyson}} &=&G_{\mu \nu }^{\left( \text{WD}%
\right) }\left( \phi \right) d\phi ^{\mu }d\phi ^{\nu }  \notag \\
&=&G_{\mu \nu }^{\left( \text{P}\right) }\left( \theta \left( \phi \right)
\right) d\theta \left( \phi \right) ^{\mu }d\theta \left( \phi \right) ^{\nu
}=\frac{4}{\phi ^{2}}d\phi ^{2}\text{.}  \label{gg}
\end{eqnarray}%
Equations (\ref{P_line}) and \textbf{(}\ref{gg}) will be used in our IGAC in
quantum spin chains systems. Before considering such information-geometric
characterization of quantum energy level statistics for regular and chaotic
spin chains immersed in an external magnetic field, we briefly review the
main points of the more standard approach to these topics.

\subsection{Entanglement in quantum spin chains: standard formalism}

One of the most important concepts in quantum information theory is that of
entanglement, an intrinsic property of composite quantum systems.
Entanglement plays an essential role in many-body quantum phenomena, such as
superconductivity \cite{tinkham} and quantum phase transitions \cite{sachdev}%
. Moreover, it is an important concept in quantum computation and
information processing \cite{nielsen}. An excellent theoretical framework
for investigating entanglement properties is offered by spin chains. Quantum
spin chains belong to the most studied models of quantum statistical
mechanics. However, only for a few types of models have the thermal and
ground state structures have been determined. This is mainly a consequence
of the complicated correlations that can arise among quantum states. These
strong correlations can even be present in pure quantum states, while
classical pure states can only have a trivial product state structure.
Unlike the classical case, the restrictions of pure states on the quantum
spin chain to local subsystems are typically mixed states. This type of
correlation between subsystems is commonly referred to as entanglement. The
von Neumann entropy, defined as,%
\begin{equation}
\mathcal{S}_{\text{von Neumann}}=-tr\left( \rho \log \rho \right) \text{,}
\end{equation}%
is a standard measure of the nonpurity of the reduced density matrix $\rho $%
, thus it is a very useful quantity in the description of entanglement \cite%
{bennett}. Several simple models of spin chains can be studied analytically
and there also exist efficient numerical techniques. There are two widely
used methods of characterizing entanglement in spin chains. The first of
these describes the entanglement between two spins in the chain with a
quantity called concurrence \cite{osborne}. The other one measures
entanglement of a block of spins with the rest of the chain with the von
Neumann entropy when the chain is in its ground state \cite{keating}. As a
side remark, we emphasize that entanglement entropy does not refer
exclusively to the characterization of quantum systems in their ground
states, but, more generally, it refers to any many particles quantum
dynamical states undergoing a unitary time evolution\textbf{.} The method
used in \cite{keating} is known as the density matrix renormalization group
method (DMRG) \cite{white}. It is based on the fact that many degrees of
freedom are redundant in quantum state description; therefore, the system is
adequately described by taking into account maximally entangled components
only. von Neumann entropy is supposed to play an important role in
quantifying the essential subspace of a reduced density matrix. The
possibility of compressing such density matrices from its full dimension to
a much smaller subspace without significant loss of information is the
starting point of the DMRG analysis. Classical complexity of quantum states
can be characterized by a mixed state entanglement entropy. For instance,
the von Neumann entropy of a block of $L$ neighboring spins in a $XX$ chain,
describing entanglement of the block with the rest of the chain is given by 
\cite{eisler},%
\begin{equation}
\mathcal{S}_{L}=-tr\left( \rho _{L}\log \rho _{L}\right) \overset{%
L\rightarrow \infty }{\propto }\log L\text{.}
\end{equation}%
The reduced density matrix $\rho _{L}$\ is obtained from the ground state $%
\left\vert \Psi _{g}\right\rangle $\ of the chain by tracing out external
degrees of freedom,%
\begin{equation}
\rho _{L}=tr_{N-L}\left\vert \Psi _{g}\right\rangle \left\langle \Psi
_{g}\right\vert \text{, }\mathcal{H}\left\vert \Psi _{g}\right\rangle
=E_{g}\left\vert \Psi _{g}\right\rangle \text{.}  \label{sos}
\end{equation}%
The Hamiltonian $\mathcal{H}$ in (\ref{sos})\ is given by \cite{keating,
eisler},%
\begin{equation}
\mathcal{H}=-\overset{N}{\underset{l=1}{\sum }}\left(
s_{l}^{x}s_{l+1}^{x}+s_{l}^{y}s_{l+1}^{y}\right) -h\overset{N}{\underset{l=1}%
{\sum }}s_{l}^{z}\text{,}
\end{equation}%
where $s_{l}^{\alpha }\left( \alpha =x\text{, }y\text{, }z\right) $\ are the
Pauli spin matrices at sites $l=1$, $2$,.., $N$\ of a periodic chain and $h$%
\ is the magnetic\textbf{\ }field. The logarithmic growth of the
entanglement entropy is a general consequence of the fact that in one
dimensional systems near quantum phase transitions, the entropy is a
logarithmic function of the size of the system \cite{calabrese}.
Furthermore, there is also a time dependent version of this DMRG, known as $%
\tau $-DMRG \cite{white(2)}. This method is used to study the evolution of
pure states, density matrices and operators. Note that the classical
complexity of quantum operators can be characterized using the operator
space entanglement entropy of a density operator, not the state entanglement
entropy of a mixed state. For the evolution of density\textbf{\ }matrices
and operators, a superket corresponding to an operator $\mathcal{O}$
expanded in the computational basis of products of local operators is
considered. For instance, for a chain of $n$-qubits, a basis of $4^{n}$\
Pauli operators is used, $\sigma ^{s_{0}}\otimes ...\otimes \sigma
^{s_{n-1}} $, with $s_{j}\in \left\{ 0\text{, }x\text{, }y\text{, }z\right\} 
$\ and $\sigma ^{0}=I$. The key idea of $\tau $-DMRG is to represent any
operator in a matrix product form \cite{prosen(pre)},%
\begin{equation}
\mathcal{O}=\underset{s_{j}}{\sum }tr\left(
A_{0}^{s_{0}}...A_{n-1}^{s_{n-1}}\right) \sigma ^{s_{0}}\otimes ....\otimes
\sigma ^{s_{n-1}}
\end{equation}%
in terms of the $4n$\ matrices $A_{j}^{s_{j}}$\ of fixed dimension $D$. The
number of parameters in the matrix product (MPO) representation of the
operator is $4nD^{2}$. The minimal $D$\ required, $D_{\varepsilon }\left(
\tau \right) $, is equal to the maximal rank of the reduced super density
matrix over bipartitions of the chain. The way entropy can be computed in
the spaces of operators can be found in \cite{prosen(pre)}. The $\tau $-DMRG
method is very efficient in classical simulations of many body quantum
dynamics requiring that the computational costs grow polynomially in time
and, consequently, that the entanglement entropy grows no faster than
logarithmically. However, it is known that the asymptotic behavior of
computational costs and entanglement entropies of integrable and chaotic
Ising spin chains are very different \cite{prosen(07)}. Here Prosen
considered the question of time efficiency implementing an up-to-date
version of the $\tau $-DMRG for a family of Ising spin $\frac{1}{2}$ chains
in arbitrary oriented magnetic field, which undergoes a transition from
integrable (transverse Ising) to nonintegrable chaotic regime as the
magnetic field is varied. An integrable (regular) Ising chain in a general
homogeneous transverse magnetic field is defined through the Hamiltonian $%
\mathcal{H}^{\left( \text{regular}\right) }\equiv \mathcal{H}\left( 0\text{, 
}2\right) $, where%
\begin{equation}
\mathcal{H}\left( h_{x}\text{, }h_{y}\right) =\underset{j=0}{\overset{n-2}{%
\sum }}\sigma _{j}^{x}\sigma _{j+1}^{x}+\underset{j=0}{\overset{n-1}{\sum }}%
\left( h^{x}\sigma _{j}^{x}+h^{y}\sigma _{j}^{y}\right) \text{.}
\label{ising-hamiltonian}
\end{equation}%
In this case, the computational cost shows a polynomial growth in time, $%
D_{\varepsilon }^{\left( \text{regular}\right) }\left( \tau \right) \overset{%
\tau \rightarrow \infty }{\propto }\tau $, while the entanglement entropy is
characterized by logarithmic growth,%
\begin{equation}
\mathcal{S}^{\left( \text{regular}\right) }\overset{\tau \rightarrow \infty }%
{\propto }c\log \tau +c^{\prime }\text{. }  \label{regularentropy}
\end{equation}%
The constant\textbf{\ }$c$\textbf{\ }depends exclusively on the value of the
fixed transverse magnetic field intensity\textbf{\ }$B_{\perp }$, while%
\textbf{\ }$c^{\prime }$\textbf{\ }depends on\textbf{\ }$B_{\perp }$\textbf{%
\ }and on the choice of the initial local operators of finite index used to
calculate the operator space\textbf{\ }entanglement entropy. Instead, a
quantum chaotic Ising chain in a general homogeneous tilted magnetic field
is defined through the Hamiltonian $\mathcal{H}^{\left( \text{chaotic}%
\right) }\equiv \mathcal{H}\left( 1\text{, }1\right) $, where $\mathcal{H}$
is defined in (\ref{ising-hamiltonian}). In this case, the computational
cost shows an exponential growth in time, $D_{\varepsilon }^{\left( \text{%
chaotic}\right) }\left( \tau \right) \overset{\tau \rightarrow \infty }{%
\propto }\exp \left( \mathcal{K}_{q}\tau \right) $ while the entanglement
entropy is characterized by linear growth,%
\begin{equation}
\mathcal{S}^{\left( \text{chaotic}\right) }=\overset{\tau \rightarrow \infty 
}{\propto }\mathcal{K}_{q}\tau \text{.}
\end{equation}%
The quantity $\mathcal{K}_{q}$\ is a constant, asymptotically independent of
the number of indexes of the initial local operators used to calculate the
operator space entropy, that depends only on the Hamiltonian evolution and
not on the details of the initial state observable or error measures, and
can be interpreted as a kind of quantum dynamical entropy.

It is well known the quantum description of chaos is characterized by a
radical change in the statistics of quantum energy levels \cite{casati and
company}. The transition to chaos in the classical case is associated with a
drastic change in the statistics of the nearest-neighbor spacings of quantum
energy levels. In the regular regime the distribution agrees with the
Poisson statistics while in the chaotic regime the Wigner-Dyson distribution
works very well. Uncorrelated energy levels are characteristic of quantum
systems corresponding to a classically regular motion while a level
repulsion (a suppression of small energy level spacing) is typical for
systems which are classically chaotic. A standard quantum example is
provided by the study of energy level statistics of an Hydrogen atom in a
strong magnetic field. It is known that level spacing distribution (LSD) is
a standard indicator of quantum chaos \cite{haake}. It displays
characteristic level repulsion for strongly nonintegrable quantum systems,
whereas for integrable systems there is no repulsion due to existence of
conservation laws and quantum numbers. In \cite{prosen(07)}, the authors
calculate the LSD of the spectra of\textbf{\ }$\mathcal{H}^{\left( \text{%
regular}\right) }$ and $\mathcal{H}^{\left( \text{chaotic}\right) }$. They
find that for $\mathcal{H}^{\left( \text{regular}\right) }$, the nearest
neighbor LSD is described by a Poisson distribution. For $\mathcal{H}%
^{\left( \text{chaotic}\right) }$, they find the nearest neighbor LSD is
described by a Wigner-Dyson distribution. Therefore, they conclude that $%
\mathcal{H}^{\left( \text{regular}\right) }$ and $\mathcal{H}^{\left( \text{%
chaotic}\right) }$ indeed represent generic regular and quantum chaotic
systems, respectively.

In the next paragraph, we will encode the relevant information about the
spin-chain in a suitable composite-probability distribution taking into
account the quantum spin chains and the configurations of the external
magnetic field in which they are immersed.

\section{An information Geometric Model of Regular and Chaotic Quantum Spin
Chains}

\subsection{Integrable Statistical Model: Poisson coupled to an Exponential
Bath}

Recall that in the ME\ method \cite{caticha2}, the selection of relevant
variables is made on the basis of intuition guided by experiment; it is
essentially a matter of trial and error. The variables should include those
that can be controlled or experimentally observed, but there are cases where
others must also be considered. Our objective here is to choose the relevant
microvariables of the system and select the relevant information concerning
each one of them. In the integrable case, the Hamiltonian $\mathcal{H}%
^{\left( \text{regular}\right) }$ describes an antiferromagnetic Ising chain
immersed in a transverse homogeneous magnetic field $\vec{B}_{\text{%
transverse}}=B_{\perp }$ $\hat{B}_{_{\perp }}$ with the level spacing
distribution of its spectrum given by the Poisson distribution%
\begin{equation}
p_{\text{A}}^{\left( \text{Poisson}\right) }\left( x_{\text{A}}|\mu _{\text{A%
}}\right) =\frac{1}{\mu _{\text{A}}}\exp \left( -\frac{x_{\text{A}}}{\mu _{%
\text{A}}}\right) \text{,}  \label{poisson-integrable}
\end{equation}%
where the microvariable $x_{\text{A}}$ is the spacing of the energy levels
and the macrovariable $\mu _{\text{A}}$ is the average spacing. The chain is
immersed in the transverse magnetic field which has just one component $%
B_{\perp }$ in the Hamiltonian $\mathcal{H}^{\left( \text{regular}\right) }$%
. We translate this piece of information in our IGA formalism, coupling the
probability (\ref{poisson-integrable}) to an exponential bath $p_{\text{B}%
}^{\left( \text{exponential}\right) }\left( x_{\text{B}}|\mu _{\text{B}%
}\right) $ given by%
\begin{equation}
p_{\text{B}}^{\left( \text{exponential}\right) }\left( x_{\text{B}}|\mu _{%
\text{B}}\right) =\frac{1}{\mu _{\text{B}}}\exp \left( -\frac{x_{\text{B}}}{%
\mu _{\text{B}}}\right) \text{,}
\end{equation}%
where the microvariable $x_{\text{B}}$ is the intensity of the magnetic
field and the macrovariable $\mu _{\text{B }}$is the average intensity. More
correctly, $x_{\text{B}}$\ should be the energy arising from the interaction
of the magnetic field with the spin $\frac{1}{2}$\ particle magnetic moment, 
$x_{\text{B}}=\left\vert -\vec{\mu}\cdot \vec{B}\right\vert =\left\vert -\mu
B\cos \varphi \right\vert $\ where $\varphi $\ is the tilt angle. For the
sake of simplicity, let us set $\mu =1$, then in the transverse case $%
\varphi =0$\ and therefore $x_{\text{B}}=B\equiv B_{\perp }$. This is our
best guess and we justify it by noticing that the magnetic field intensity
is indeed a relevant quantity in this experiment (see equation (\ref%
{regularentropy})) and its components (intensity) are quantities that are
varied during the transitions from integrable to chaotic regimes. In the
regular regime, we say the magnetic field intensity is set to a well-defined
value $\left\langle x_{\text{B}}\right\rangle =\mu _{\text{B}}$.
Furthermore, notice that the exponential distribution is identified by
information theory as the maximum entropy distribution if only one piece of
information (the expectation value) is known. Finally, the chosen composite
probability distribution $P^{\left( \text{integrable}\right) }\left( x_{%
\text{A}}\text{, }x_{\text{B}}|\mu _{\text{A}}\text{, }\mu _{\text{B}%
}\right) $ encoding relevant information about the system is given by,%
\begin{eqnarray}
P^{\left( \text{integrable}\right) }\left( x_{\text{A}}\text{, }x_{\text{B}%
}|\mu _{\text{A}}\text{, }\mu _{\text{B}}\right) &=&p_{\text{A}}^{\left( 
\text{Poisson}\right) }\left( x_{\text{A}}|\mu _{\text{A}}\right) p_{\text{B}%
}^{\left( \text{exponential}\right) }\left( x_{\text{B}}|\mu _{\text{B}%
}\right)  \notag \\
&&  \notag \\
&=&\frac{1}{\mu _{\text{A}}\mu _{\text{B}}}\exp \left[ -\left( \frac{x_{%
\text{A}}}{\mu _{\text{A}}}+\frac{x_{\text{B}}}{\mu _{\text{B}}}\right) %
\right] \text{.}  \label{p-exp}
\end{eqnarray}%
Again, we point out that our probability (\ref{p-exp}) is our best guess
and, of course, must be consistent with numerical simulations and
experimental data in order to have some merit. We point out that equation (%
\ref{p-exp}) is not fully justified from a theoretical point of view, a
situation that occurs due to the lack of a systematic way to select the
relevant microvariables of the system (and to choose the appropriate
information about such microvariables). Let us denote $\mathcal{M}%
_{S}^{\left( \text{integrable}\right) }$ the two-dimensional curved
statistical manifold underlying our information geometrodynamics. The line
element $\left( ds^{2}\right) _{\text{integrable}}$ on $\mathcal{M}%
_{S}^{\left( \text{integrable}\right) }$ is given by,%
\begin{equation}
\left( ds^{2}\right) _{\text{integrable}}=\frac{1}{\mu _{\text{A}}^{2}}d\mu
_{\text{A}}^{2}+\frac{1}{\mu _{\text{B}}^{2}}d\mu _{\text{B}}^{2}\text{.}
\label{yes}
\end{equation}%
Applying our IGAC to the line element in (\ref{yes}) and following the steps
provided in the ED\ Gaussian model of Sections II and III of this paper, we
obtain polynomial growth in $\mathcal{V}_{\mathcal{M}_{s}}^{\text{integrable}%
}$ and logarithmic IGE growth, 
\begin{equation}
\mathcal{V}_{\mathcal{M}_{s}}^{\left( \text{integrable}\right) }\left( \tau
\right) \overset{\tau \rightarrow \infty }{\propto }\exp (c_{IG}^{\prime
})\tau ^{c_{IG}}\text{, }\mathcal{S}_{\mathcal{M}_{s}}^{\left( \text{%
integrable}\right) }\left( \tau \right) \overset{\tau \rightarrow \infty }{%
\propto }c_{IG}\log \tau +c_{IG}^{\prime }\text{.}  \label{1}
\end{equation}%
The quantity\textbf{\ }$c_{IG}$\textbf{\ }is a constant proportional to the
number of exponential probability distributions in the composite
distribution used to calculate the IGE and\textbf{\ }$c_{IG}^{\prime }$%
\textbf{\ }is a constant that depends on the values assumed by the
statistical macrovariables\textbf{\ }$\mu _{A}$\textbf{\ }and\textbf{\ }$\mu
_{B}$. Equations (\ref{1}) may be interpreted as the information-geometric
analogue of the computational complexity $D_{\varepsilon }^{\left( \text{%
regular}\right) }\left( \tau \right) $\ and the entanglement entropy $%
\mathcal{S}^{\left( \text{regular}\right) }$\ defined in standard quantum
information theory, respectively. We cannot state they are the same since we
are not fully justifying, from a theoretical standpoint, our choice of the
composite probability (\ref{p-exp}).

\subsection{Chaotic Statistical Model: Wigner-Dyson coupled to a Gaussian
Bath}

In the chaotic case, the Hamiltonian $\mathcal{H}^{\left( \text{chaotic}%
\right) }$ describes an antiferromagnetic Ising chain immersed in a tilted
homogeneous magnetic field $\vec{B}_{\text{tilted}}=B_{\perp }$ $\hat{B}%
_{\perp }+B_{\parallel }$ $\hat{B}_{\parallel }$ with the level spacing
distribution of its spectrum given by the Poisson distribution $p_{\text{A}%
}^{\left( \text{Wigner-Dyson}\right) }\left( x_{\text{A}}^{\prime }|\mu _{%
\text{A}}^{\prime }\right) $%
\begin{equation}
p_{\text{A}}^{\left( \text{Wigner-Dyson}\right) }\left( x_{\text{A}}^{\prime
}|\mu _{\text{A}}^{\prime }\right) =\frac{\pi x_{\text{A}}^{\prime }}{2\mu _{%
\text{A}}^{\prime 2}}\exp \left( -\frac{\pi x_{\text{A}}^{\prime 2}}{4\mu _{%
\text{A}}^{\prime 2}}\right) \text{.}  \label{wigner-dyson(chaotic)}
\end{equation}%
where the microvariable $x_{\text{A}}^{\prime }$ is the spacing of the
energy levels and the macrovariable $\mu _{\text{A}}^{\prime }$is the
average spacing. The chain is immersed in the tilted magnetic vector field
which has two components $B_{\perp }$ and $B_{\parallel }$ in the
Hamiltonian $\mathcal{H}^{\left( \text{chaotic}\right) }$. We translate this
piece of information in our IGAC formalism, coupling the probability (\ref%
{wigner-dyson(chaotic)}) to a Gaussian $p_{\text{B}}^{\left( \text{Gaussian}%
\right) }\left( x_{\text{B}}^{\prime }|\mu _{\text{B}}^{\prime }\text{, }%
\sigma _{\text{B}}^{\prime }\right) $ given by,%
\begin{equation}
p_{\text{B}}^{\left( \text{Gaussian}\right) }\left( x_{\text{B}}^{\prime
}|\mu _{\text{B}}^{\prime }\text{, }\sigma _{\text{B}}^{\prime }\right) =%
\frac{1}{\sqrt{2\pi \sigma _{\text{B}}^{\prime 2}}}\exp \left( -\frac{\left(
x_{\text{B}}^{\prime }-\mu _{\text{B}}^{\prime }\right) ^{2}}{2\sigma _{%
\text{B}}^{\prime 2}}\right) \text{.}
\end{equation}%
where the microvariable $x_{\text{B}}^{\prime }$ is the intensity of the
magnetic field, the macrovariable $\mu _{\text{B }}^{\prime }$is the average
intensity, and $\sigma _{\text{B}}^{\prime }$ is its covariance:\ during the
transition from the integrable to the chaotic regime, the magnetic field
intensity is being varied (experimentally). It is being tilted and its two
components ($B_{\perp }$ and $B_{\parallel }$) are being varied as well. Our
best guess based on the experimental mechanism that drives the transitions
between the two regimes is that magnetic field intensity ( actually the
microvariable $\mu B\cos \varphi $) is Gaussian-distributed (two
macrovariables) during this change. In the chaotic regime, we say the
magnetic field intensity is set to a well-defined value $\left\langle x_{%
\text{B}}^{\prime }\right\rangle =\mu _{\text{B}}^{\prime }$\ with
covariance $\sigma _{\text{B}}=\sqrt{\left\langle \left( x_{\text{B}%
}^{\prime }-\left\langle x_{\text{B}}^{\prime }\right\rangle \right)
^{2}\right\rangle }$. Furthermore,\textbf{\ }the Gaussian distribution is
identified by information theory as the maximum entropy distribution if only
the expectation value and the variance are known. Therefore, the chosen
composite probability distribution $P^{\left( \text{chaotic}\right) }\left(
x_{\text{A}}^{\prime }\text{, }x_{\text{B}}^{\prime }|\mu _{\text{A}%
}^{\prime }\text{, }\mu _{\text{B}}^{\prime }\text{, }\sigma _{\text{B}%
}^{\prime }\text{ }\right) $ encoding relevant information about the system
is given by,%
\begin{eqnarray}
P^{\left( \text{chaotic}\right) }\left( x_{\text{A}}^{\prime }\text{, }x_{%
\text{B}}^{\prime }|\mu _{\text{A}}^{\prime }\text{, }\mu _{\text{B}%
}^{\prime }\text{, }\sigma _{\text{B}}^{\prime }\text{ }\right) &=&p_{\text{A%
}}^{\left( \text{Wigner-Dyson}\right) }\left( x_{\text{A}}^{\prime }|\mu _{%
\text{A}}^{\prime }\right) p_{\text{B}}^{\left( \text{Gaussian}\right)
}\left( x_{\text{B}}^{\prime }|\mu _{\text{B}}^{\prime }\text{, }\sigma _{%
\text{B}}^{\prime }\right)  \notag \\
&&  \notag \\
&=&\frac{\pi \left( 2\pi \sigma _{\text{B}}^{\prime 2}\right) ^{-\frac{1}{2}}%
}{2\mu _{\text{A}}^{\prime 2}}x_{\text{A}}^{\prime }\exp \left[ -\left( 
\frac{\pi x_{\text{A}}^{\prime 2}}{4\mu _{\text{A}}^{\prime 2}}+\frac{\left(
x_{\text{B}}^{\prime }-\mu _{\text{B}}^{\prime }\right) ^{2}}{2\sigma _{%
\text{B}}^{\prime 2}}\right) \right] \text{.}
\end{eqnarray}%
Let us denote $\mathcal{M}_{S}^{\left( \text{chaotic}\right) }$ the
three-dimensional curved statistical manifold underlying our information
geometrodynamics. The line element $\left( ds^{2}\right) _{\text{chaotic}}$
on $\mathcal{M}_{S}^{\left( \text{chaotic}\right) }$ is given by,%
\begin{equation}
\left( ds^{2}\right) _{\text{chaotic}}=\frac{4}{\mu _{\text{A}}^{\prime 2}}%
d\mu _{\text{chain}}^{2}+\frac{1}{\sigma _{\text{B}}^{\prime 2}}d\mu _{\text{%
B}}^{\prime 2}+\frac{2}{\sigma _{\text{B}}^{\prime 2}}d\sigma _{\text{B}%
}^{\prime 2}\text{.}  \label{yes2}
\end{equation}%
Applying our IGAC to the line element in (\ref{yes2}) and following the
steps provided in the ED\ Gaussian model of Sections II and III of this
paper, we obtain exponential growth in $\mathcal{V}_{\mathcal{M}%
_{s}}^{\left( \text{chaotic}\right) }$ and linear IGE growth,%
\begin{equation}
\mathcal{V}_{\mathcal{M}_{s}}^{\left( \text{chaotic}\right) }\left( \tau
\right) \overset{\tau \rightarrow \infty }{\propto }C_{IG}\exp \left( 
\mathcal{K}_{IG}\tau \right) \text{, }\mathcal{S}_{\mathcal{M}_{s}}^{\left( 
\text{chaotic}\right) }\left( \tau \right) \overset{\tau \rightarrow \infty }%
{\propto }\mathcal{K}_{IG}\tau \text{.}  \label{yu}
\end{equation}%
The constant\textbf{\ }$C_{IG}$\textbf{\ }encodes information about the
initial conditions of the statistical macrovariables parametrizing elements
of $\mathcal{M}_{S}^{\left( \text{chaotic}\right) }$. The constant $\mathcal{%
K}_{IG}$,%
\begin{equation}
\mathcal{K}_{IG}\overset{\tau \rightarrow \infty }{\approx }\frac{d\mathcal{S%
}_{\mathcal{M}_{s}}\left( \tau \right) }{d\tau }\overset{\tau \rightarrow
\infty }{\approx }\underset{\tau \rightarrow \infty }{\lim }\left[ \frac{1}{%
\tau }\log \left( \left\Vert \frac{J_{\mathcal{M}_{S}}\left( \tau \right) }{%
J_{\mathcal{M}_{S}}\left( 0\right) }\right\Vert \right) \right] \overset{%
\text{def}}{=}\lambda _{J}\text{,}
\end{equation}%
is the model parameter of the chaotic system and depends on the temporal
evolution of the statistical macrovariables. It plays the role of the
standard Lyapunov exponent of a trajectory and it is, in principle, an
experimentally observable quantity. The quantity\textbf{\ }$J_{\mathcal{M}%
_{S}}\left( \tau \right) $\textbf{\ }is the Jacobi field intensity and%
\textbf{\ }$\lambda _{J}$\textbf{\ }may be considered the
information-geometric analogue of the leading Lyapunov exponent in
conventional Hamiltonian systems. Given an explicit expression of\textbf{\ }$%
\mathcal{K}_{IG}$\textbf{\ }in terms of the observables\textbf{\ }$\mu _{%
\text{A }}^{\prime }$\textbf{\ }and\textbf{\ }$\mu _{\text{B }}^{\prime }$%
\textbf{\ }and\textbf{\ }$\sigma _{\text{B}}^{\prime }$, a clear
understanding of the relation between the IGE (or\textbf{\ }$\mathcal{K}%
_{IG} $) and the entanglement entropy (or $\mathcal{K}_{q}$) becomes the key
point that deserves further study.\textit{\ }Equations (\ref{yu}) are the
information-geometric analogue of the computational complexity $%
D_{\varepsilon }^{\left( \text{chaotic}\right) }\left( \tau \right) $ and
the entanglement entropy $\mathcal{S}^{\left( \text{chaotic}\right) }$
defined in standard quantum information theory, respectively. This result is
remarkable, but deserves a deeper analysis in order to be fully understood.

One of the major limitations of our findings is the lack of a detailed
account of the comparison of the theory with experiment. This point will be
one of our primary concerns in future works. However, some considerations
may be carried out at the present stage. The experimental observables in our
theoretical models are the statistical macrovariables characterizing the
composite probability distributions. In the integrable case, where the
coupling between a Poisson distribution and an exponential one is considered,%
\textbf{\ }$\mu _{\text{A }}$\textbf{\ }and\textbf{\ }$\mu _{\text{B }}$%
\textbf{\ }are the experimental observables. In the chaotic case, where the
coupling between a Wigner-Dyson distribution and a Gaussian is considered,%
\textbf{\ }$\mu _{\text{A }}^{\prime }$\ and\textbf{\ }$\mu _{\text{B }%
}^{\prime }$\textbf{\ }and $\sigma _{\text{B}}^{\prime }$\textbf{\ }play the
role of the experimental observables. We believe one way to test our theory
may be that of determining a numerical estimate of the leading Lyapunov
exponent\textbf{\ }$\lambda _{\text{max}}$\textbf{\ }or the Lyapunov
spectrum for the Hamiltonian systems under investigation directly from
experimental data (measurement of a time series) and compare it to our
theoretical estimate for\textbf{\ }$\lambda _{J}$\textbf{\ }\cite{wolf}.
However, we are aware that it may be extremely hard to evaluate Lyapunov
exponents numerically. Otherwise, knowing that the mean values of the
positive Lyapunov exponents are related to the Kolmogorov-Sinai (KS)
dynamical entropy, we suggest to measure the KS entropy\textbf{\ }$\mathcal{K%
}$\textbf{\ }directly from a time signal associated to a suitable
combination of our experimental observables and compare it to our indirect
theoretical estimate for\textbf{\ }$\mathcal{K}_{IG}$\textbf{\ }from the
asymptotic behaviors of our statistical macrovariables \cite{procaccia}. We
are aware that the ground of our discussion is quite qualitative. However,
we hope that with additional study, especially in clarifying the relation
between the IGE and the entanglement entropy, our theoretical
characterization presented in this paper will find experimental support in
the future. Therefore, the statement that our findings may be relevant to
experiments verifying the existence of chaoticity and related dynamical
properties on a macroscopic level in energy level statistics in chaotic and
regular quantum spin chains is purely a conjecture at this stage.

\section{FINAL REMARKS}

In this paper, we reviewed our novel information-geometrodynamical approach
to chaos (IGAC) on curved statistical manifolds and we emphasized the
usefulness of our information-geometrodynamical entropy (IGE) as an
indicator of chaoticity in a simple application. Furthermore, knowing that
integrable and chaotic quantum antiferromagnetic Ising chains are
characterized by asymptotic logarithmic and linear growths of their operator
space entanglement entropies, respectively, we applied our IGAC to present
an alternative characterization of such systems. Remarkably, we have shown
that in the former case the IGE exhibits asymptotic logarithmic growth while
in the latter case the IGE exhibits asymptotic linear growth.

It is worthwhile emphasizing the following points:\ the statements that
spectral correlations of classically integrable systems are well described
by Poisson statistics and that quantum spectra of classically chaotic
systems are universally correlated according to Wigner-Dyson statistics are
conjectures, known as the BGS (Bohigas-Giannoni-Schmit, \cite{bohigas} and
BTG (Berry-Tabor-Gutzwiller, \cite{gutzwiller}) conjectures, respectively.
These two conjectures are very important in the study of quantum chaos,
however their validity finds some exceptions. Several other cases may be
considered. For instance, chaotic systems having a spectrum that does not
obey a Wigner-Dyson distribution may be considered. A chaotic system can
also have a spectrum following a Poisson, semi-Poisson, or other types of
critical statistics \cite{garcia}. Moreover, integrable systems having a
spectrum that does not obey a Poisson distribution may be considered as
well. For instance, the Harper model would represent such a situation.
Moreover, it is worthwhile pointing out that not every chaotic system
characterized by entropy-like quantities growing linearly in time has a
spectrum described by a Wigner-Dyson distribution. Well-known examples
presenting such a situation are the cat maps \cite{gu} and the famous kicked
rotator \cite{izrailev} where its spectrum follows a Poisson distribution in
cylinder representation and a Wigner-Dyson in torus representation but the
properties of entropy-like quantities are the same in both representations
(at least classically). All these cases are not discussed in our
characterization.

Therefore, at present stage, because of the above considerations and because
of the lack of experimental evidence in support of our theoretical
construct, we can only conclude that the IGAC might find some potential
applications in certain regular and chaotic dynamical systems and this
remains only a conjecture. However, we hope that our work convincingly shows
that this information-geometric approach may be considered a serious effort
trying to provide a unifying criterion of chaos of both classical and
quantum varieties, thus deserving further research.

\begin{acknowledgments}
The authors are grateful to Prof. Ariel Caticha and Dr. Adom Giffin for
useful comments. We thank an anonymous Referee for constructive criticism
that lead to concrete improvement of this work.
\end{acknowledgments}

\end{document}